\documentclass[12pt]{JHEP3}
\pdfoutput=1
\usepackage{amsmath,epsfig}
\usepackage{amssymb,amsfonts}
\usepackage{latexsym}
\usepackage[latin1]{inputenc}
\usepackage{float}

\usepackage{subeqnarray}
\usepackage{xcolor}

\usepackage{graphicx}
\usepackage{longtable}
\usepackage{enumerate}

\relax
\renewcommand{\theequation}{\arabic{section}.\arabic{equation}}
\def\be{\begin{equation}}
\def\ee{\end{equation}}
\def\bea{\begin{eqnarray}}
\def\eea{\end{eqnarray}}
\newcommand\fverb{\setbox\pippobox=\hbox\bgroup\verb}
\newcommand\fverbdo{\egroup\medskip\noindent%
                        \fbox{\unhbox\pippobox}\ }
\newcommand\fverbit{\egroup\item[\fbox{\unhbox\pippobox}]}

\newcommand{\bear}{\begin{eqnarray}}

\newcommand{\eear}{\end{eqnarray}}

\newcommand{\bsea}{\begin{subeqnarray}}
\newcommand{\esea}{\end{subeqnarray}}
\newbox\pippobox

\newcommand{\ga}{\gamma}

\newcommand{\ud}{\mathrm{d}}

\def\6{\partial}

\def\a{\alpha}
\def\g{\gamma}

\def\half{\frac12}

\def\le{\left}
\def\ri{\right}

\def\e{\epsilon}
\def\m{\mu}
\def\n{\nu}
\def\r{\rho}
\def\s{\sigma}

\def\sp{,\;\;}

\def\z{\zeta}
\def\sq
\def\a{\alpha}
\def\b{\beta}
\def\l{\lambda}
\def\y{\psi}

\def\hri#1#2{\href{http://arxiv.org/abs/#1}{[ArXiv:#1]#2}}

\def\h{\eta}
\def\e{\epsilon}

\def\d{\delta}

\def\L{\Lambda}

\title{Magnetic Critical Solutions in Holography}
\author{{\large N. Angelinos\\
 ~\\
 \href{http://hep.physics.uoc.gr/}
 {Crete Center for Theoretical Physics}, Department of Physics, University of Crete
 71003 Heraklion, Greece}}

\preprint{CCTP-2014-21}

\abstract{The AdS/CFT correspondence is a realization of the holographic principle in the context of string theory. It is a map between a quantum field theory and a string theory living in one or more extra dimensions. Holography provides new tools to study strongly-coupled quantum field theories. It has important applications in quantum chromodynamics (QCD) and condensed matter (CM) systems, which are usually complicated and strongly coupled. 
Quantum critical CM theories have scaling symmetries and can be connected to higher-dimensional scale invariant space-times. The Effective Holographic Theory paradigm may be used to describe the low-energy (IR) holographic dynamics of quantum critical systems by the Einstein-Maxwell-Dilaton (EMD) theory.
We find the magnetic critical scaling solutions of an EMD theory containing an extra parity-odd term $F\wedge F$. Previous studies in the absence of magnetic fields have shown the existence of quantum critical lines separated by quantum critical points. We find this is also true in the presence of a magnetic field. The critical solutions are characterized by the triplet of critical exponents ($\theta,z,\zeta$), the first two describing the geometry, while the latter describes the charge density. }

 %%%%%%%TeX, LaTeX, % %%%%%%%%NesTeX} %%%%%%%%%\dedicated{Dedicated to\ldots\\if you want.}

\begin{document}

\section{Introduction}
\label{intro}

The AdS/CFT correspondence, also known as gauge/gravity correspondence, suggests a duality between a string theory and a quantum field theory. In its earliest incarnation, \cite{Mald} the gauge/gravity duality connected a 10-dimensional theory on anti-de Sitter space-time (AdS$_5\times S^5$) with a 4-dimensional conformal field theory (CFT) living on its boundary. It was discovered in the context of string theory by studying field theories on hypersurfaces (D-branes) embedded in a higher dimensional space-time, and the associated black holes.

A strongly coupled field theory corresponds to a weakly coupled string theory, which is easier to handle. Because of that, holography provides an alternative way of studying strongly coupled QFTs. The AdS/CFT correspondence has a wide variety of applications ranging from quantum chromodynamics (AdS/QCD) to condensed matter physics (AdS/CMT) and to relativistic hydrodynamics.

In this paper we focus on quantum critical theories at finite charge density. Such theories can be strongly coupled and hard to study in more than 1 dimensions. However, quantum critical points have special symmetries and the theory can be connected to a simple gravitational dual. We study quantum critical solutions at finite density in the presence of a magnetic field, using the holographic approach.

This thesis is structured as follows. We first motivate and introduce the AdS/CFT correspondence, \cite{Mald} and explain how it realizes holography. We then introduce the Effective Holographic Theories, \cite{eht} which are useful in the study of quantum critical points. Finally we discuss our setup and solutions. Details about the calculations can be found in the appendices.

\section{The AdS/CFT correspondence and Holography}

In this section we present the basic idea of the Holographic Principle. We also discuss the connection between large-N gauge theories with string theories. We explain how the AdS/CFT correspondence arises when studying a system of D-branes inside a 10-dimensional bulk space-time, \cite{Mald} and we present some of its applications.

\subsection{What is Holography}
The Holographic Principle states that a theory with gravity in a closed region of space-time is described by degrees of freedom that live on the boundary. The description of the volume is completely encoded on a dual theory living on its boundary.

Holography was inspired by Bekenstein's bound, which states that in a theory of gravity the maximal entropy in a region of space scales with its surface area, rather than its volume. This condition is saturated by black holes. Bekenstein argued that if this bound was not satisfied by a system, then it would be possible to violate the second law of thermodynamics, \cite{bek}. In contrast, local quantum field theory predicts that the number of degrees of freedom inside a region scales with its volume. For that reason, Bekenstein's bound has been controversial and is one of the points that makes QFT seemingly incompatible with gravity. It is believed, however, that a successful theory of quantum gravity must satisfy the Holographic bound. Holography has been embedded in the framework of string theory and is a widely studied subject.

The most succesful realization of Holography is the Anti de-Sitter/Conformal Field Theory (AdS/CFT) correspondence discovered by J. Maldacena in 1997, \cite{Mald}. A superstring theory on AdS$_5\times S^5$ is conjectured to be equivalent to 4-dimensional $\mathcal{N}=4$ super Yang-Mills. The latter is a conformal field theory (CFT) living on the boundary of the space-time, \cite{witten}. The theories are equivalent even though they live in a different number of dimensions. Every field in the AdS string theory can be translated to an operator in the CFT and vice versa.

\subsection{Large-N Gauge Theories and String Theory}
It was first suggested by 't Hooft that a strongly coupled gauge theory in the large-N limit can be described by an effective string theory at weak coupling, \cite{hooft}. In this section we introduce some basic ideas of string theory and present the stucture of its perturbation theory, which turns out to be a topological expansion. We also present the perturbative structure of large-N Yang-Mills theory, which has a topological expansion of identical structure if we identify the string coupling constant $g_s$ with $1/N$, \cite{st}. The conclusion is that in the large-N limit the effective string theory description is weakly coupled (small $g_s$).
\subsubsection{String Theory}
We start from the simple case of a relativistic particle. Its motion describes a curve in spacetime. This curve is called the world-line of the particle. The equation of motion of this particle can be derived by minimizing the length of its world-line between two points, \cite{st}.
\be S=-m\int ds \label{a1}\ee
where m is the mass of the particle and ds the line element of its trajectory.

%We can parametrize the trajectory of the particle using a parameter $\tau$:
%\be x^\m=x^\m(\tau)\ee
%If at the starting and ending points of the trajectory the parameter takes values ($t_i$,$t_f$) respectively the action (\ref{a1}) can be written explicitly as:
%\be S=-m\int_{t_i}^{t_f} \ud\tau \sqrt{-\eta_{\m\n}\dot{x}^\m \dot{x}^\n}\ee
%where $\eta_{\m\n}$ is the Minkowski metric.

The action for a relativistic string is built following the same idea. A relativistic string is a one-dimensional continuous object and its motion in space-time describes a two-dimensional surface (world-sheet). In analogy with the relativistic particle, the equations of motion for the string are derived by minimizing the surface of its world-sheet between two string configurations. In the case of closed strings the world-sheet is a tube, while for open strings it is a strip. The action describing the motion of the string is the Nambu-Goto action, \cite{NG}, \cite{st}:
\be S_{NG}=-T\int dA \label{a2}\ee
where $T=(2\pi \ell_s^2)^{-1}$ is the tension of the string and dA is the surface element of its worldsheet. The parameter $\ell_s$ is the characteristic length of the theory (string length).

We can take two coordinates $\xi^a (a=0,1)$ to parametrize the world-sheet $\Sigma(\xi^0,\xi^1)$. The string moves in a space-time with metric $G_{\m\n}$ (target space). The target space induces a metric on the world-sheet, \cite{st}:
\be ds^2=G_{\m\n}(X) dX^\m dX^\n=G_{\m\n}{\partial X^\m\over \partial \xi^a}{\partial X^\n\over \partial \xi^b}d\xi^a d\xi^b=\tilde G_{ab} d\xi^a d\xi^b \ee
where the induced metric is
\be\tilde G_{ab}=G_{\m\n}{\partial X^\m\over \partial \xi^a}{\partial X^\n\over \partial \xi^b} \ee
The Nambu-Goto action (\ref{a2}) can be written explicitly as, \cite{st}:
\be S_{NG}=-T\int \ud^2 \xi \sqrt{-\det \tilde G_{ab}} \ee
The equations of motion can be solved in flat target space-time for both Neumann and Dirichlet boundary conditions. The relativistic string can then be quantized by following standard methods, like canonical quantization, \cite{st}.

\begin{figure}[H]
\centering
\includegraphics[width=1\textwidth]{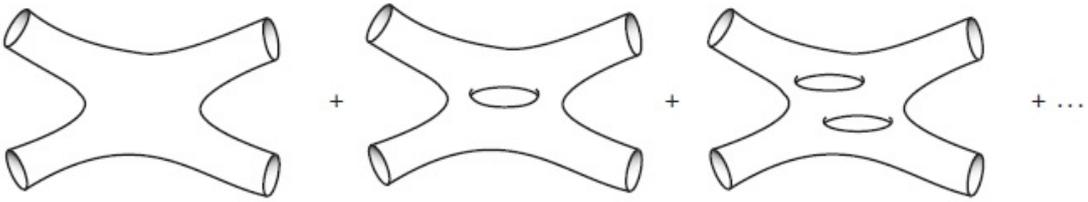}
\caption{The topological expansion of closed string theory. This figure was taken from \cite{AdS/CFT}}.
\label{strings}
\end{figure}

We are not interested in the details of string quantization for our purpose, but we examine the perturbative structure of string theory. In closed string theory the basic interaction is a string splitting into two (or the inverse), \cite{st}. The world-sheet of such an interaction can be obtained by thickening the lines of a triple vertex in QFT. A one-loop diagram can be obtained by combining two triple vertex diagrams and the resulting world-sheet is a Riemann surface with one hole (genus 1). Repeating this process we obtain higher order diagrams of higher genus. Their world-sheet is a surface characterized by the topological number, h, which is the number of holes (see figure \ref{strings}). Therefore the perturbation series in string theory is a topological expansion. By weighing each triple vertex with the dimensionless string coupling constant $g_s$ the string perturbative expansion is of the form, \cite{st}:
\be\sum_{h=0}^{\infty} g_s^{2h-2} F_h(a') \label{s1}\ee
As we will see below the perturbative expansion of large-N gauge theories is identical to (\ref{s1}).

\subsubsection{Large-N Gauge Theories}
Consider a U(N) Yang-Mills theory:
\be\mathcal{L}=-{1\over g_{YM}^2}Tr[F_{\m\n}F^{\m\n}] \label{l1} \ee
 The gauge field strength tensor is given by:
\be F_{\m\n}=\partial_\m A_\n-\partial_\n A_\m+[A_\m,A_\n]\ee
The gauge field components $A_\m$ are $N\times N$ Hermitian matrices and the theory is non-Abelian.

In order to take the large-N limit, we first have to know how to scale the coupling constant $g_{YM}$. In quantum field theory the beta function encodes the dependence of the coupling parameter $g_{YM}$ on the energy scale $\m$. From the one-loop beta function for a non-Abelian U(N) gauge theory we obtain the RG flow equation, \cite{st}:
\be \m {\ud g_{YM}\over \ud \m}=-{11\over 3}N {g_{YM}^3\over (4\pi)^2}+\mathcal{O}(g_{YM}^5)\ee
We can find the appropriate scaling by demanding that the leading terms are of the same order. Therefore we can see that the combination
\be \lambda=g_{YM}^2 N \label{tH}\ee
which is called the 't Hooft coupling, \cite{hooft}, must be kept constant as N goes to infinity.
The Lagrangian (\ref{l1}) can be rewritten as
\be\mathcal{L}=-{N\over\l}Tr[F_{\m\n}F^{\m\n}]\ee

The 't Hooft expansion corresponds to the expansion of the amplitudes in powers of N while keeping $\l$ constant. A convenient way to write the vacuum to vacuum diagrams is the 't Hooft double-line notation, \cite{hooft}, which substitutes each line in the Feynman diagrams with two lines of opposite orientations, {\cite{st}}.

\begin{figure}[H]
\centering
\includegraphics[width=0.6\textwidth]{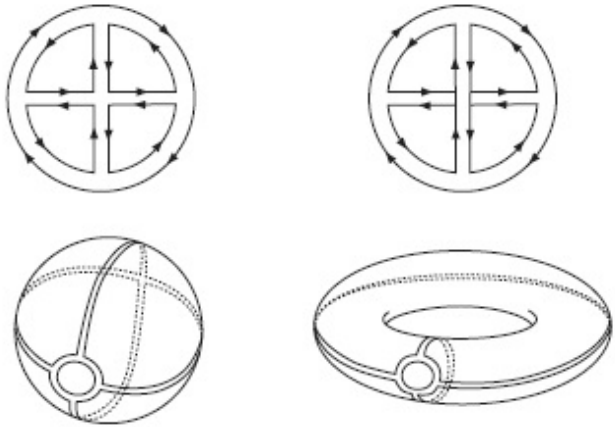}
\caption{On the left: the zeroth order diagram can be drawn on a surface with zero handles (sphere). On the right: the first order diagram cannot be drawn on a sphere, but has to be drawn on a surface of genus 1 (torus). This figure was taken from \cite{AdS/CFT}}.
\label{diagrams}
\end{figure}

Every propagator contributes a factor of $\l/N$ and every vertex a factor of $N/\l$. In addition every loop contributes a power of N (because of the summation over N colors). We can now find the factors associated with a diagram. A diagram with E propagators (edges), V vertices and
F loops (faces) has a coefficient proportional to \cite{st}:
\be \left({\l\over N}\right)^E \left({N\over \l}\right)^V N^F=N^\chi \l^{E-V}\ee
where $\chi=V-E+F$ is the Euler number of the surface, \cite{st}. For a closed compact surface with $h$ handles $\chi=2-2h$. Therefore such a diagram has a coefficient of order $\mathcal{O}(N^{2-2h})$. The 't Hooft expansion organizes diagrams according to their topology, \cite{hooft}. In the large-N limit the dominant diagrams are the ones with the minimum number of handles ($h=0$), which have the topology of a sphere. These diagrams are called planar because they can also be drawn on a plane. Non-planar diagrams correspond to surfaces with handles and are suppressed in the large-N limit by an additional factor of $1/N^{2h}$.

The standard perturbative expansion for any correlator can be written at large N as, \cite{st}:
\be  \sum_{h=0}^\infty N^{2-2h} Z_h(\l)=\sum_{h=0}^\infty N^{2-2h}\sum_{i=0}^\infty c_{i,h}\l^i\ee
which suggests a connection with the topological expansion of string theory (\ref{s1}) if we identify the string coupling constant $g_s$ with:
\be g_s\sim 1/N\ee
This connection indicates that in the large-N limit the effective string theory description is weakly coupled.

\subsection{The benchmark example: $AdS_5/CFT _4$}

String theory contains, besides strings, objects extended in more than one spatial dimensions, called branes, \cite{st}. The most important ones for the AdS/CFT correspondence are the D$_p$-branes, which are defined as (p+1)-dimensional hypersurfaces on which open strings can end with Dirichlet boundary conditions. On one hand the brane dynamics can be described pertubatively in terms of open strings. On the other hand, the D-branes interact gravitationally and are supergravity solutions. In this section we will review how the AdS/CFT correspondence was discovered by conjecturing a duality between these two different descriptions of the same system of D-branes, \cite{Mald}.

\subsubsection{Charged D-branes}
$D_p$-branes have two types of excitations, \cite{AdS/CFT}. The first type is the motion and deformation of their shapes which can be parametrized by their coordinates $\phi^i$ in the (9-p)-dimensional transverse space. These degrees of freedom are scalar fields on the brane's world-volume. The second type are internal excitations caused by the charged end of a string. A charge is a source to a gauge field and the D$_p$-brane has an Abelian gauge field A$_\m, (\m=0,1,...,p)$ living on its world-volume. The action that describes these types of excitations is the Dirac-Born-Infeld (DBI) action, \cite{AdS/CFT}:
\be S_{DBI}=-T_{D_p}\int \ud^{p+1}x\sqrt{-\det({g_{\m\n}}+2\pi \ell_s^2 F_{\m\n})}\label{DBI}\ee
where $T_{D_p}=(2\pi)^{-p}g_s^{-1}\ell_s^{-p-1}$ is the tension of the brane, $\ell_s$ the string length and $g_{\m\n}$ is the induced metric on the brane's world-volume which depends on the scalar fields $\phi^i$.

Consider now a $D_3$-brane in flat target space. We can write the induced metric on the brane using the 6 scalar fields $\phi^i$ as $g_{\m\n}=\eta_{\m\n}+(2\pi \ell_s^2)^2 \partial_\m \phi^i \partial_\n\phi^i$. The DBI action (\ref{DBI}) for a $D_3$-brane can be written as:
\be S_{brane}=-T_{D_3}\int \ud^{4}x\sqrt{-\det\left[\eta_{\m\n}+(2\pi \ell_s^2\partial_\m \phi^i) (2\pi \ell_s^2\partial_\n \phi^i)+2\pi \ell_s^2 F_{\m\n}\right]}\ee
We notice that every field $\phi, F$ is accompanied by a factor of $2\pi \ell_s^2$. We can now expand in powers of $2\pi \ell_s^2$ (which is equivalent to expanding in powers of the gravitational constant $\kappa$ and keeping the string coupling $g_s$ constant since $\kappa\sim g_s \ell_s^4$, \cite{st}, using the familiar identity from linear algebra: $\det(A)=\exp[Tr(\log A)]$, where A is a square matrix. The leading order terms are (ignoring the constant zeroth order term):
\be S_{brane}=-{1\over 2\pi g_s}\int \ud^4 x \left({1\over 4} F_{\m\n}F^{\m\n}+{1\over 2}\partial_\m \phi^i\partial_\n \phi^i+\dots\right) \label{DBI2}\ee
where we used $$T_{D_3}=((2\pi)^3 g_s \ell_s^4)^{-1}$$ The rest of the terms are suppressed by additional factors of $\ell_s^2$. Therefore at the low-energy limit we have a U(1) theory living on the world-volume of the $D_3$-brane.

We now consider a system of N coincident parallel D$_3$-branes ($p=3$). Such a system generate a non-Abelian U(N) gauge theory. The branes live in a 10-dimensional bulk space and are located at the same point of the transverse 6-dimensional space. Except for the open strings, which are excitations of the branes, the theory also contains closed strings, which are excitations of the bulk space. We can write the low-energy action of this theory schematically:
\be S=S_{bulk}+S_{brane}+S_{interactions}\ee
The first term $S_{bulk}$ describes the dynamics of the bulk space in terms of closed strings, which are described by IIB superstring theory. The second term $S_{brane}$ describe the dynamics of the branes in terms of open strings. The third term $S_{interactions}$ contains the interactions between open and closed strings.

At the low-energy limit, the superstring theory living on the bulk reduces to free IIB supergravity. By expanding S$_{bulk}$ around the free point in powers of the gravitational constant $\kappa=8\pi G_N$ as $g_{\m\n}=\h_{\m\n}+\kappa h_{\m\n}$ we obtain, \cite{st}:
\be S_{bulk}\sim {1\over 2\kappa^2}\int \ud^{10} x \sqrt{-det(g_{\m\n})}R+\dots\sim \int\ud^{10} x \left[(\partial h)^2 + \kappa h (\partial h)^2+\dots\right] \ee
where we have not indicated explicitly all the bulk fields for simplicity. The important result is that all the interaction terms are proportional to positive powers of $\kappa$, therefore at very low energies (small $\kappa$) they become very weak compared to the kinetic term and can be ignored.

The second term S$_{brane}$ governs the open degrees of freedom (open strings). As we indicated earlier, a single $D_3$-brane has a U(1) theory living on its world-volume containing a gauge field and 6 scalar fields. A system of N branes will generate a U(N) theory containing a vector boson and 6 scalars transforming as adjoints of U(N). In particular such a system is equivalent to $\mathcal{N}=4, U(N)$ super Yang-Mills (SYM) theory in the low-energy limit, \cite{st}:
\be S_{brane}\sim-{1\over 2\pi g_s}\int \ud^4 x  Tr\left[{1\over 4}F_{\m\n}F^{\m\n}\right]+\dots \label{DBI3}\ee
where we kept only the gauge field terms for simplicity.

The third term describing the interactions between open-closed degrees of freedom is subleading in the low energy limit. We expand S$_{interactions}$ in powers of $\kappa$, \cite{st}:
\be S_{interactions}\sim \int \ud^4 x\sqrt{-det(g_{\m\n})} Tr\left[F^2\right]+\dots\sim \kappa \int \ud^4 x h_{\m\n} Tr\left[F_{\m\n}^2-{\delta_{\m\n}\over 4}F^2\right]+\dots \ee
where again the only terms indicated are the kinetic terms of the gauge field for simplicity.

We conclude that in the low-energy limit, this theory is described by free IIB supergravity on the bulk and $\mathcal{N}=4, U(N)$ super Yang-Mills theory on the branes not interacting with each other.

\subsubsection{D-branes as supergravity solutions}
We now consider the same system from a different point of view. The $D$-branes are solutions of supergravity in 10-dimensions. The exact solution for N $D_3$-branes is given by, \cite{st}:
\be ds^2=H^{-1/2}(-dt^2+d\vec{x}^2)+H^{1/2}(dr^2+r^2 d\Omega_5^2)\label{m1}\ee
The 3 spatial coordinates $\vec{x}$ are parallel to the branes, while $dr^2+r^2 d\Omega_5^2$ is the metric of the 6-dimensional transverse space. In particular $r$ is the distance from the branes, and the metric changes as we move along $r$ because of the warp factor:
\be H=1+{L^4\over r^4}\sp L^4=4\pi g_s l_s^4 N\label{w}\ee

Since the coefficient of $dt^2$ in the metric depends on the distance from the branes, $r$, the energy measured also depends on $r$ (due to gravitational redshift). If at some point $r$ we measure energy $E_r$, an observer at infinity would measure $E_\infty=H^{-{1\over 4}}E_r$, \cite{st}. Therefore an object moving near the branes $r\to 0$ (which means $H^{-{1\over 4}}\to 0$) would appear to have very low energy to an observer at infinity.

From the point of view of an observer located at infinity there are two types of low energy excitations: massless low energy modes that propagate in the whole bulk and all modes that approach the horizon ($r=0$), since any finite energy is redshifted to zero. In the low-energy limit the two types of excitations decouple from each other.

Excitations of the first kind propagate away from the branes where the space is flat. For very large r, $H\approx 1$ and the metric (\ref{m1}) becomes flat. Therefore this set of modes is described by free supergravity.

Excitations of the second kind remain confined near the horizon, because there is a potential barrier they have to climb in order to get away, \cite{st}. For very small $r$, $H^{1/2}\approx L^2/r^2$ and the metric (\ref{m1}) becomes:
\be ds^2={r^2\over L^2}(-dt^2+d\vec{x}^2)+{L^2\over r^2}(dr^2+r^2 d\Omega_5^2)\ee
Changing the radial coordinate to $u=L^2/r$ we obtain:
\be ds^2={L^2\over r^2}(du^2-dt^2+d\vec{x}^2)+L^2 d\Omega_5^2 \label{AdSS}\ee
which describes the product space $AdS_5\times S^5$. Therefore the low-energy limit of this system is described by IIB free supergravity and IIB supergravity in $AdS_5\times S^5$ which are not interacting with each other.

In both descriptions we notice that the low-energy description of the system of N $D_3$-branes reduces to the sum of two non-interacting theories. In both cases one of those theories is free IIB supergravity. We therefore expect that the two remaining theories, gravity in $AdS_5\times S^5$ and $\mathcal{N}=4, U(N)$ SYM, are equivalent. The latter is a conformal field theory (CFT) in 4-dimensions. This is why this special case of holography is called $AdS_5/CFT_4$ correspondence.

\subsubsection{Validity of AdS/CFT correspondence}
We now examine the region of validity of the two dual descriptions. We would first like to find the connection between the dimensionless parameters of the two theories. Starting from (\ref{DBI3}) and keeping only the gauge field terms:
\be \mathcal{L}_{YM}={1\over 2\pi g_s}  Tr\left[{1\over 4}F_{\m\n}F^{\m\n}\right]={c\over 2\pi g_s} {1\over 4} F^a_{\m\n}F^{a,\m\n}\sp F_{\m\n}=F^a_{\m\n}T^a\ee
where $T^a$ are the generators of the non-Abelian group and c is their normalization constant: $Tr\left[T^a T^b\right]=c \delta^{ab}$. A popular choice that we are going to use is $c=1/2$, \cite{st}. Therefore the Yang-Mills coupling is:
\be g_{YM}^2=4\pi g_s \label{gYM}\ee
Combining (\ref{gYM}) with the expression for the radius of the AdS space-time from (\ref{w}):
\be \left({L\over \ell_s}\right)^4=N g_{YM}^2=\lambda \label{f1}\ee
where $\lambda$ is the 't Hooft coupling defined in (\ref{tH}). This is one of the formulas we were looking for.

The 10-dimensional Newton constant is given by:
\be 16\pi G_{10}=(2\pi)^7 g_s^2 \ell_s^8\ee
Combining this with (\ref{gYM}) and (\ref{tH}) we obtain:
\be {G_{10}\over \ell_s^8}={\pi^4 \lambda^2\over 2 N^2}\label{g10}\ee
We can now find the region of validity of the two descriptions.
From (\ref{g10}) we see that $G_{10}\sim 1/N^2$, which means that quantum effects are suppressed for large N. If the CFT is strongly coupled ($\lambda\gg 1$) then, according to (\ref{f1}), $L\gg \ell_s$, which means that the string theory is weakly curved and can be approximated by supergravity. Therefore the large-N limit of the sYM theory is described well by the two-derivative action of IIB supergravity in $AdS_5\times S^5$.

\subsubsection{Symmetries}

We will now briefly compare the symmetries of the two descriptions. We first consider the conformal symmetry. The $\mathcal{N}=4$ SYM theory is a 4-dimensional CFT. Therefore it is invariant under the conformal group SO(2,4). This is exactly the isometry group of the $AdS_5\times S^5$ background of the dual string theory, \cite{st}.

We now consider the sypersymmetries of the two descriptions. The $\mathcal{N}=4$ SYM theory is maximally supersymmetric with 32 conserved fermionic supercharges. The $AdS_5\times S^5$ is also a maximally supersymmetric solution of 10-dimensional supergravity with 32 Killing spinors, which correspond to the 32 supercharges of the dual gauge theory, \cite{st}. In addition $\mathcal{N}=4$ SYM is invariant under the R-symmetry group SO(6), which rotates the 6 scalar fields $\phi^i$ into each other. This symmetry can be identified with the rotational symmetry of the 5-sphere component of the dual $AdS_5\times S^5$ space-time, \cite{st}.

\subsubsection{The extra dimension as an energy scale}

Consider the limit on the energy as we are taking $\ell_s\to 0$ in the brane theory. The energy an observer at infinity measures is:
\be E_\infty=H^{-{1\over 4}}E_r\sim E_r{r\over \ell_s} \ee
We must keep the energy in the near-horizon region fixed in string units $\ell_s E_r$, \cite{st}, as well as the energy in the near-boundary region $E_\infty$ since this is the energy measured in the CFT. From:
\be E_\infty\sim E_r{r\over l_s}=(E_r \ell_s){r\over \ell_s^2}\ee
we see that $U={r\over \ell_s^2}$ must be kept fixed as we are taking the $r\to 0$ limit. The radial coordinate is therefore proportional to the energy scale of the dual CFT near the horizon. We can change the radial coordinate to $U={r\over \ell_s^2}$ in the near-horizon metric, \cite{Mald}:
\be ds^2=\ell_s^2 \left[{U^2\over \sqrt{4\pi g_s N}}(-dt^2 + d\vec{x}^2)+ \sqrt{4\pi g_s N}\left({dU^2\over U^2}+d\Omega_5^2\right)\right]\ee
The coordinates $t,\vec{x}$ are the space-time coordinates of the CFT. The extra coordinate, $U$, of the AdS part behaves like the energy scale of the CFT.

This argument involved the decoupling limit. There are, however, other arguments that indicate that the radial direction behaves like an energy scale for the gauge theory with the UV located near the boundary. We write the $AdS_5$ metric in Poincar\'e coordinates:
\be ds^2={1\over u^2}(du^2-dt^2+d\vec{x}^2)\ee
The metric is invariant under SO(1,1):
\be (u,t,\vec{x})\to (au,at,a\vec{x})\ee
The boundary is located at $u=0$ in these coordinates. If we scale up the coordinates of the gauge theory on the boundary $(t,\vec{x})$, which means going down in energy, we must also scale up u, which means moving away from the boundary at $u=0$. Therefore the high-energy limit of the gauge theory (UV) corresponds to $u = 0$ (the boundary) while the low-energy limit (IR) corresponds to $u = \infty$ (the Poincar\'e horizon), \cite{st}.

\subsubsection{Counting degrees of Freedom}
In this section we introduce a cutoff in the $AdS_5$ metric and calculate the degrees of freedom (entropy) of the theory on the boundary and the bulk theory.

We begin by cutting out the 5-sphere part of the metric (\ref{AdSS}). It turns out that by reducing the $S^5$ part, every massless field in the original theory corresponds to an infinite tower of massive fields on AdS$_5$ with ever increasing mass, \cite{st}, \cite{AdS/CFT}. The only interesting detail for our purpose is the value of the 5-dimensional Newton's constant $G_5$ after the reduction on $S^5$. The relationship between $G_{10}$ and $G_5$ can be found by considering the reduction of the Einstein-Hilbert term, \cite{AdS/CFT}:
\be {1\over 16\pi G_{10}} \int \ud^5 x \ud ^5 \Omega \sqrt{-\det(g_{10})}R_{10}={V\over 16\pi G_{10}} \int \ud^5 x \sqrt{-\det(g_{5})}R_{5}\ee
where $V=\pi^3 L^5$ is the volume of an $S^5$ of radius L. It follows that
\be G_5=G_{10}/V={(2\pi)^7l_s^8 g_s^2\over (16\pi) \pi^3 L^5}={\pi L^3\over 2 N^2}\ee
where we also used (\ref{w}).

Now that we removed the sphere we consider the AdS$_5$ metric in global coordinates (more details can be found in appendix \ref{AdS}):
\be ds^2=L^2(-\cosh^2(\rho)d\tau^2+d\rho^2 +\sinh^2(\rho)d\Omega_3^2) \label{m5}\ee
Changing the radial coordinate to $u=\tanh({\rho\over 2})$, (\ref{m5}) becomes:
\be ds^2=L^2(-\left({1+u^2\over 1-u^2}\right)^2d\tau^2+{4\over (1-u^2)^2}(du^2+u^2 d\Omega_3^2)) \label{m4}\ee

In this coordinate system the boundary is located at $u=1$ and the interior at $0\leq u<1$.

We introduce a cutoff close to the boundary at $u^2=1-\e$ with $\e$ very small. This corresponds to a UV cutoff in the dual gauge theory according to the previous section. The gauge theory lives on an $S^3$ of radius L. The distance between two points on the cutoff sphere scales as $\log(|x_1-x_2|/\epsilon)$. Therefore we may view $\e\ll 1$ as a small distance cutoff in the gauge theory. Since the gauge theory has a small distance cutoff $L\e$ (in units of length), there are $1/\e^3$ fundamental cells of radius L in the 3-sphere. Since the gauge theory has order $N^2$ degrees of freedom and the entropy is proportional to the regulated volume of the boundary we obtain:
\be S_{FT}\sim {N^2\over \e^3}\ee

On the $AdS_5$ side the area of the sphere at the regulated boundary can be read directly from the metric (\ref{m4}):
\be A= 8  {L^3 u^3\over (1-u^2)^3}\bigg|_{u^2=1-\e}\sim {L^3\over \e^3}\label{a}\ee
The gravitational entropy is given by the Bekenstein bound:
\be S_{AdS}\sim{A\over G_5}\sim {N^2\over \e^3}\ee
where $G_5$ is the AdS$_5$ Newton constant we calculated earlier. The two descriptions have the same degrees of freedom. Therefore the AdS/CFT correspondence successfully realized holography in string theory.

We can also calculate the volume of $AdS_5$ up to the shifted boundary $u^2=1-\e$:
\be V_4=L^4 \int {2u^3\over (1-u^2)^4}\ud u \ud^3\Omega_3=L^4 \Omega_3 \int_{0}^{u^2=1-\e} {2u^3\over (1-u^2)^4}\ud u\sim {L^4\over \e^3}\ee
where $\Omega_3$ is the volume of the 3-sphere. Comparing with (\ref{a}) we notice that the volume of $AdS_5$ scales with the same power of $\e$ as its area close to the boundary. From that aspect holography seems trivial. However the non-triviality of our previous analysis stems from the fact that area and volume scale with different powers of L.

\subsection{Applications}
The gauge/gravity correspondence has a wide range of applications, mostly because of the fact that it enables the study of a strongly coupled field theory in terms of a weakly coupled theory with gravity. We briefly discuss the most important areas where the AdS/CFT correspondence is applied.
\begin{itemize}

\item{\bf AdS/QCD}

Quantum chromodynamics, the theory of strong interactions, becomes strongly coupled at low energies. Holography provides a new method to study the low-energy spectrum of the theory, \cite{QCD,QCD0}. There are two approaches to AdS/QCD. The first is the top-down approach, in which a brane configuration in string theory is engineered so that its low-energy spectrum shows poperties similar to QCD, \cite{QCD1,QCD2}. The second is the bottom-up approach, where models are built based on phenomenology, \cite{QCD3,QCD4,QCD5,QCD6,QCD7,QCD8,QCD9,QCD10}.

\item{\bf AdS/CMT}

Theories describing interesting states of matter such as superconductors, superfluids, Bose-Einstein condensates as well as theories describing transitions at zero temperature (quantum critical points) can be strongly coupled. Most of these phenomena are well studied by experiments, however they may be difficult to explain theoretically using the usual techniques from QFT. In holography, superconductors and superfluids are studied at finite temperature in terms of a black hole in a higher dimensional space-time. It has been discovered that near the horizon of an AdS black hole the Abelian symmetry can be spontaneously broken \cite{ADSCMT}, \cite{Hartnoll}, which is analogous to what happens when a system transits to a superconducting phase. There has been some success in the holographic description of superconductors, superfluids and insulators, as well as superfluid/insulator and superconductor/insulator transitions \cite{eht,SFI,SCI,SC1,SC2,SC3,SC4,SC5,SC6,SC7,SC8,SC9,SC10,SC11,SC12,SC13,SC14,SC15}.

\item{\bf Fluid/gravity correspondence}

Hydrodynamics is an effective long-distance description of any QFT that is locally in thermodynamic equilibrium.
Fluid dynamics is described by the Navier-Stokes equations, which are non-linear partial differential equations and are difficult to solve. Fluid dynamics can be connected to a classical gravitational theory, \cite{FG1,FG2,FG3,FG4,FG5,FG6,FG7,FG8,FG9,FG10,FG11,FG12,FG13,FG14}. The geometries dual to fluid dynamics are black hole space-times with
regular event horizons. It has also been discovered that the cosmological Einstein equations reduce to the relativistic Navier-Stokes equations at the appropriate long-wavelength limit and it has been shown that there is a one to one correspondence between dissipative Navier-Stokes equations Einstein equations coupled to matter. Results are summarized in \cite{FG3,FG4,FGC}.

\end{itemize}

\newpage
\section{Effective Holographic Theories}

The concept of Effective Holographic Theories, \cite{eht}, is very useful in the study of holographic IR fixed points. They are developed in analogy with Effective Field Theories (EFT) which are used to study the low-energy limit of a QFT. In analogy with EFTs, the strategy is to select a collection of fields that dominates the low-energy dynamics and build an action that governs their behavior. However a string theory, in general, has an infinite number of fields with non-zero vacuum expectation values. The central point of the approximation is to truncate string theory to a finite spectrum, keeping only a few fields dual to the most important QFT operators.

\subsection{Building the EHT}

The main idea of EHTs is to select a collection of operators that are expected to dominate the low-energy dynamics and build a general action containing their dual string fields at the two-derivative level.

We first consider the simplest case. A quantum field theory always has a stress-energy tensor, therefore the EHT must contain a spin-2 tensor, the metric, which encodes the energy distribution of the dual field theory. The minimal theory containing a metric with AdS solution is the Einstein gravity with a cosmological constant.

In a system at finite charge density another important operator is the conserved current $J_\m$. In this case we include a massless gauge field $A_\m$ in the holographic theory, dual to the conserved current. The theory in this case is the Einstein-Maxwell theory. Using these two fields the basic solution is the AdS-Reissner-N\"ordstrom black hole, \cite{RN1}, \cite{RN2} which has many interesting properties.

The next step is to add a scalar field dual to the most important scalar operator. The resulting EHT is an Einstein-Maxwell-Dilaton (EMD) theory. This theory has 3 fields, the metric $g_{\m\n}$ which controls the energy distribution of the field theory, the gauge field $A_\m$ which controls the charge density, and a scalar $\phi$ controlling a scalar coupling constant and vacuum expectation value (vev).

\subsection{Einstein-Maxwell-Dilaton theory \label{EMD}}

The most general action at the two-derivative level containing a scalar, a U(1) gauge field and a metric is the EMD theory, which is given in (d+1)-dimensions by:
\be S_{EMD}=M^{d-1}\int \ud^{d+1}x\le[ \sqrt{-g}\left(R-\frac12(\partial \phi)^2+V(\phi)-{1\over 4}Z(\phi) F^2\right)\ri] \ee
In the above action $d$ is the number of the space-time dimensions of the dual field theory. The action contains the Ricci scalar, $R$, a scalar potential $V$ and the gauge field strength tensor $F_{\m\n}=\partial_\m A_\n -\partial_\n A_\m$.

The asymptotic behavior of the coupling functions when $\phi\to \pm\infty$ is motivated by generic examples in string theory and is exponential. We parametrize it as:
\be V(\phi)=e^{-\d\phi}\sp Z(\phi)=e^{\g\phi}\label{e1}\ee

Zero temperature solutions to this theory have geometries with scaling symmetries that should be dual to quantum critical theories. We will analyze this topic in the next section.

Zero temperature solutions, in general, have naked singularities. Although in classical gravity such solutions are unacceptable, in holography they are not always unphysical. Gubser has studied and presented the criterion for acceptable singularities in holography \cite{Singularities}. We expect that singularities that satisfy the Gubser criterion are resolvable. This can happen by either embedding them in a higher-dimension solution (see for example \cite{gk2}) or by stringy effects \cite{dab}.

\subsection{The story so far\label{ssf}}

In this section we describe the most important results from previous studies of the EMD theory. The zero temperature scaling solutions of the EMD theory have a metric of the form:
\be ds^2=r^\theta \left(-{dt^2\over r^{2z}}+{dr^2+dx_1^2+dx_2^2\over r^2}\right)\ee
This metric has the following scaling symmetry:
\be t\to \l^z t\sp r\to \l r\sp x^i\to \l x^i\sp ds\to\l^{\theta/2}ds\ee
This symmetry appears in CM theories near quantum critical points.

The gauge field in general scales in the IR as
\be A_t=\m+Qr^{\zeta-z}\ee
where $\m,Q$ are identified as the chemical potential and charge density of the field theory.

A solution is characterized by the 3 critical exponents ($z,\theta,\zeta$).
The exponent $z$ is the {\em dynamical critical exponent}, or {\em Lifshitz exponent} and $\theta$ is the {\em hyperscaling violation exponent}. CFTs have $\theta=0\sp z=1$, while Lifshitz theories have $\theta=0\sp z\ne 1$. The {\em conductivity exponent}, $\zeta$, is in general independent of the other two exponents and determines the behavior of the charge density in the IR.

When $\theta\ne 0$ the proper distance $ds$ also has to be scaled. This indicates violation of hyperscaling in the boundary theory, \cite{HSS}. In (d+1)-dimensional field theories with hyperscaling symmetry the free energy of the system scales by its naive dimension. In theories with Lorentz symmetry ($z=1\sp \theta=0$) the entropy is proportional to $\sim T^d$, where $d$ is the number of spatial dimensions of the field theory. In Lifshitz theories ($z\ne 1\sp \theta=0$) the entropy S depends on the temperature as $S\sim T^{d/z}$. When hyperscaling is violated ($z\ne 1\sp \theta\ne 0$) the entropy scales as $S\sim T^{d-\theta\over z}$ and the field theory has an effective dimensionality of $d_{eff}=d-\theta$.

The behavior of the exponents $z,\theta$ is closely related to the behavior of the charge density and scalar field respectively. The solutions are scale invariant ($z=1$) in the absence of charges, while inhomogeneous metrics ($z\ne 1$) can be obtained when the charge density is finite to leading order. The hyperscaling violating exponent $\theta$ depends on the behavior of the scalar. We obtain hyperscaling violating geometries ($\theta\ne 0$) by letting the scalar run logarithmically to $\pm\infty$ in the IR.

Recent studies of EMD in the absence of magnetic fields, \cite{gk}, have shown the existence of hyperscaling violating quantum critical lines separated by hyperscaling invariant ($\theta=0$) quantum critical points. We find that this is also true in the presence of magnetic fields.

Magnetic critical solutions have been studied in the context of Einstein-Maxwell-Chern-Simons theory in \cite{magn}. The results include a magnetic zero charge density AdS$_3\times R^2$ geometry and a magnetic solution at finite density in which the charge density depends on the magnetic field. Our results include a pure magnetic AdS$_2\times R^2$ solution and a magnetic solution at finite density in which the charge density depends on the magnetic field. 

Critical solutions at finite density in the presence of an external magnetic field have been studied in \cite{gout}. Magnetic solutions of the Einstein-Maxwell-axion-dilaton theory have also been studied recently in \cite{emad}.

\section{Setup and Equations of Motion}

We consider the EMDPQ class of Effective Holographic Theories (EHT) involving the metric $g_{\m\n}$, a scalar field $\phi$ and a massless gauge field $A_\m$ in 3+1-dimensional space-time:
\be \label{action}S=M^{2}\int \ud^{4}x\le[ \sqrt{-g}\left(R-\frac12(\partial \phi)^2+V(\phi)-{1\over 4}Z(\phi) F^2-{1\over 4} W(\phi)F\wedge F\right)   \ri]\ee
with $$ F\wedge F={1\over 2\sqrt{-g}}F_{\m\n}\e^{\m\n\r\s}F_{\r\s}$$

The dual QFT in this case lives in 2+1 dimensions. The PQ (Peccei-Quinn) term, $F\wedge F$, is important when studying magnetic fields at finite density. When the magnetic field is zero it can be ignored. This term does not depend on the metric and therefore does not appear in the Einstein equations, but affects the geometry implicitly via the two other fields $\phi\sp A_\m$.

The equations of motion stemming from (\ref{action}) by varying the metric, the gauge field and scalar are respectively:
\bsea 	R_{\m\n}-{1\over 2}g_{\m\n}R&=& T_{\m\n}\slabel{EE}  \\ 	 \nabla_\mu\le(Z(\phi)F^{\mu\nu}+{W(\phi)} ^\star F^{\m\n}\ri)&=& 0\slabel{gfe}
 \\ 	 \square{\phi}+{\ud V_{eff}\over \ud \phi}\slabel{sfe}&=& 0	
\esea
where the stress energy tensor in (\ref{EE}) is given by:
\be  T_{\m\n}={V(\phi)\over 2}g_{\mu\nu}+\half\partial_\mu\phi\partial_\nu\phi-\frac{g_{\mu\nu}}4\le(\partial\phi\ri)^2+{Z(\phi)\over
2}\le[F^{\;\rho}_\mu\ F_{\nu\rho}-\frac{g_{\mu\nu}}4 F^2\ri]\ee
and the effective potential in (\ref{sfe}) is given by
\be V_{eff}(\phi)=V(\phi)-\frac{Z(\phi)}4F^2-\frac{W(\phi)}{4}F\wedge F \label{Veff}\ee

We use the standard radial ansatz for the metric and the scalar field:
\be ds^2=-D(r)dt^2+B(r)dr^2+C(r)(dx_1^2+dx_2^2)\sp \phi=\phi(r)\label{ra}\ee
The coordinates $(x_1,x_2,t)$ are the space and time coordinates of the dual $(2+1)$-dimensional field theory, while r is the holographic coordinate. This metric ansatz comes with a gauge freedom related to the freedom in reparametrizing the radial coordinate.

We also consider an electric field potential depending only on r, related to the charge density of the boundary theory. In addition, we include a uniform magnetic field in the radial direction which is related to the magnetic field of the boundary theory:
\be A_\m=(A_t(r),0,{h\over 2}x_2,-{h\over 2}x_1)\ee
The r-dependence of the magnetic field is forbidden by the Einstein equations, unless the metric is non-diagonal.
The gauge field equation (\ref{gfe}) can be integrated obtaining a conserved charge, $q$:
\be q=\sqrt{-g}Z(\phi)F^{01}-W(\phi)h\sp F^{01}={A_t'\over BD}\sp \partial_r q=0\label{q}\ee
The equations of motion can be written in the above coordinate system as:

\be \begin{split}
& \phi'^2+2{C''\over C} =\left({D'\over D}+{B'\over B}+{C'\over C}\right){C'\over C}\\
& {B\over C^2}\left({(q+hW)^2 \over
Z^2}+h^2\right)Z={D''\over D}-{C''\over C}+{1\over 2}\left({C'\over C}-{D'\over D}\right)\left({B'\over B}+{D'\over D}\right)\\
& BV+{1\over 4}{B'\over B}{C'\over C}= {1\over 2}\left({D''\over D}+{C''\over C}\right)-{1\over 4}{D'\over D}
  \left({B'\over B}+{D'\over D}-3{C'\over C}\right)\\
& \phi''+\phi'({C'\over C}+{D'\over 2D}-{B'\over 2B})={B\over 2C^2}\partial_\phi \left({(q+hW)^2\over Z}+h^2Z-2C^2V\right)\\
& A_t'={(q+hW)\over Z}{ \sqrt{DB}\over C}
\end{split}\label{EQs1}\ee
The electric and magnetic fluxes are calculated respectively by
\be \Phi_E={1\over 4\pi}\int_{R^2} ( Z(\phi) ^\star F+{W(\phi)} F)= -{V_2\over 4\pi}\left({CZA_t'\over \sqrt{BD}}-hW\right)= -{V_2\over 4\pi}q\label{EF}\ee
\be \Phi_B={1\over 4\pi}\int_{R^2} F=-{V_2\over 4\pi}h\label{MF}\ee

The integrals are over the spatial coordinates $(x_1,x_2)$ and $V_2=\int_{R^2} dx_1dx_2$ is the area of the surface of integration. The value of the electric flux depends only on the integration constant $q$ (\ref{q}). For the rest of this paper we will refer to $q$ as the electric flux and to $h$ as the magnetic field.

\subsection{Classification}

We characterize a solution as ``electric" if the electric flux (\ref{EF}) is non-zero, while the magnetic flux (\ref{MF}) vanishes in the IR. A solution is named ``magnetic" if the inverse is true. If both fluxes are non-zero we characterize the solution as ``
dyonic". We also distinguish between solutions with finite charge density ($A_t'\ne 0$) and zero charge density ($A_t'= 0$).

We split the solutions into two sections depending on the behavior of the scalar field. From (\ref{sfe}) we distinguish 2 cases:
\begin{itemize}
\item{\bf The scalar field settles to a finite constant $\phi_\star$ in the IR, which extremizes $V_{eff}$ (\ref{Veff})}

These solutions are studied in section \ref{CS}.
In the neutral case looking for asymptotic solutions of the form
\be ds^2=r^{\theta-2}(-dt^2+dr^2+dx_idx^i)\ee
we obtain an AdS$_4$ geometry. Including an electric or magnetic field the scalar equation (\ref{sfe}) is satisfied only if $C(r)$ in (\ref{ra}) is constant. The solution in the IR is AdS$_2\times R^2$. The solution can be generalized to an AdS$_2$ black hole. We study these cases in subsection \ref{geo}.
\item{\bf The scalar field runs to infinity in the IR}

These solutions are studied in section \ref{SS}. We obtain hyperscaling violating geometries by allowing the scalar to run logarithmically ($\phi=\phi_0+a\log r$) in the IR. In supergravity the coupling functions, in general, are given by a combination of exponentials of the scalar $\phi$. When $\phi$ runs to $\pm\infty$ we can keep only the exponential with the largest value. Therefore we adopt the following asymptotic behavior for the scalar coupling functions
\be V(\phi)=V_0 e^{-\d\phi}\sp Z(\phi)=Z_0e^{\g\phi}\sp W(\phi)=W_0 e^{\chi\phi} \label{cf}\ee

We look for asymptotic solutions in the deep IR, which is located either at $r\to 0$ or $r\to\infty$. For that purpose a power ansatz is enough. In particular we search for hyperscaling violating solutions which feature both a hyperscaling ($\theta$) and Lifshitz ($z$) exponent:
\be ds^2=r^{\theta}\left(-{dt^2\over r^{2z}}+{dr^2+dx_1^2+dx_2^2\over r^2}\right)\label{hv}\ee
While for the gauge field we also have the conductivity exponent $\zeta$:
\be A_t=\m+A_0 r^{\zeta-z}\ee
The constant $A_0$ is fixed by the equations and can be identified with the charge density of the boundary theory. The constant $\m$ is the chemical potential, which cannot be fixed by the equations due to the gauge invariance of the vector field.

The equations of motion using the above ansatz are:
\be
\begin{split} & (\theta-2)(\theta-2z+2) = a^2 \\
 & 2(1-z)(\theta-z-2) = B_0 r^{4-\theta}\left(h^2Z+{(q+hW)^2\over Z}\right)\\
 & (\theta-z-2)(\theta-z-1) = B_0 V r^\theta\\
 & A_0(\zeta-z)r^{\zeta-2} = \sqrt{B_0}{(q+hW)\over Z}
\end{split}\label{EQS}
\ee
Every term in the equations of motion is a power of r. Since we are looking for solutions in the deep IR, which is located either at $r\to 0$ or $r\to \infty$, we can ignore the terms that are subleading in that limit. The strategy is to solve the above equations to leading order by matching the powers and coefficients of $r$ on the left and right hand side. We distinguish cases depending on which terms are leading. More details are presented in appendix \ref{appC}.

The values of $z,\theta$ are restricted by the null-energy condition (appendix \ref{NEC}) and the allowed values are plotted in the same appendix.
The solutions are also approximately valid for small finite temperatures, at which the dual field theory has effectively $d_{eff}=2-\theta$ dimensions and the thermal entropy scales as
\be S\sim T^{2-\theta\over z}\ee
More details about hyperscaling violating metrics can be found in appendix \ref{HVM}.

Anisotropic scale invariant geometries ($z\ne 1$) are obtained when the gauge field participates to leading order in the equations. When $z=1$ the charge density is always zero in the IR.

\end{itemize}

\subsection{Electromagnetic duality of the EMD theory}
In this subsection we discuss the electromagnetic duality of the EMD theory. We rewrite the equations of motion (\ref{EQs1}):

\be \begin{split}
& \phi'^2+2{C''\over C} =\left({D'\over D}+{B'\over B}+{C'\over C}\right){C'\over C}\\
& {B\over C^2}\left({(q+hW)^2 \over
Z}+h^2Z\right)={D''\over D}-{C''\over C}+{1\over 2}\left({C'\over C}-{D'\over D}\right)\left({B'\over B}+{D'\over D}\right)\\
& BV+{1\over 4}{B'\over B}{C'\over C}= {1\over 2}\left({D''\over D}+{C''\over C}\right)-{1\over 4}{D'\over D}
  \left({B'\over B}+{D'\over D}-3{C'\over C}\right)\\
& \phi''+\phi'({C'\over C}+{D'\over 2D}-{B'\over 2B})={B\over 2C^2}\partial_\phi \left({(q+hW)^2\over Z}+h^2Z-2C^2V\right)\\
& A_t'={q+hW\over Z}{ \sqrt{DB}\over C}
\end{split}\ee

Consider the Einstein and scalar equations and ignore the gauge equation for now. The electric and magnetic fluxes $q,h$ as well as the functions $Z$ and $W$ always appear as the combination:
$$ {(q+hW)^2\over Z}+h^2Z$$
When $W$ is constant, if we swap $(q+hW)^2$ with $h^2$ and transform $Z\to 1/Z$ the equations do not change. The electromagnetic duality has the following form:
\be q\to h(1-W^2)-qW\sp h\to q+hW\sp Z\to 1/Z \label{dual}\ee
When $Z\to\infty$ the solution is purely magnetic, as the term ${(q+hW)^2\over Z}$ is subleading. When $Z\to 0$ the term $h^2Z$ is subleading and the solution is, in general, dyonic. This transformation, therefore, connects a purely magnetic solution to a dyonic one and vice versa.

Now consider the gauge field equation. When $Z\to 0$ the leading order solution is dyonic and the charge density $A_t'$ is, in general, non-zero. The electromagnetically dual solution is purely magnetic, with zero charge density $A_t'=0$. Dyonic solutions are connected to the magnetic solutions by the electromagnetic duality, with the difference that the latter exist at zero charge density.

In the special case $W=0$ the electromagnetic duality takes the simpler form:
\be q\to h\sp h\to q\sp Z\to 1/Z\ee
which maps a purely electric solution at finite charge density to a purely magnetic solution at zero charge density and vice versa. In the running scalar case we adopt the behavior (\ref{cf}) for the coupling functions and the duality takes the form $q^2/Z_0\leftrightarrow h^2Z_0\sp \g\leftrightarrow -\g$.

\newpage
\section{Constant Scalar Solutions\label{CS}}

In this section we study the solutions with the scalar settling to a finite value $\phi=\phi_\star$, that extremizes the effective potential:

\be {\ud V_{eff}\over \ud \phi}\bigg|_{\phi=\phi_\star}=0\sp V_{eff}(\phi)= V(\phi)-\frac{Z(\phi)}4F^2-\frac{W(\phi)}{4}F\wedge F\ee

The equations and details about the calculations are given in appendix \ref{appB}, while the perturbations around these solutions are presented in appendix \ref{Perturbations}. There are magnetic phases both at zero and non-zero charge density on AdS$_2\times R^2$ space-time. There are instabilities coming from the scalar field, depending on the values of the second derivatives of the coupling functions.

For the rest of this section we use the following conventions:
We define $F_\star=F(\phi_\star)$, where $F(\phi)$ is an arbitrary function of $\phi$. We also define $F'_\star={\partial\over \partial\phi}F(\phi)\bigg|_{\phi=\phi_\star}$.

\subsection{Neutral solution on AdS$_4$}\label{CS1}

We first consider the simplest case, in which both the electric and magnetic fluxes are zero ($q=0=h$). In this case the scalar settles to a constant $\phi_\star$ which extremizes the value of the scalar potential V.
\be V_\star'=0\ee
This case has been recently studied in \cite{gk}.
The leading order solution corresponds to a pure Einstein theory with a cosmological constant. Since there is no charge density the solution must be scale invariant ($z=1$). Indeed, when the value of $V_\star$ is non-zero we have an AdS$_4$ geometry:

\be ds^2={L^2\over r^2}(-dt^2+dr^2+dx_idx^i)\sp L^2={6\over V_\star}\ee

With the IR located at $r\to\infty$. Turning on the gauge field in this background we obtain
\be A_t=\m+A_1 r\ee
The second term creates a constant non-zero electric flux in the IR. It is a relevant deformation. There is also a pair of modes coming from the scalar field perturbations (see appendix \ref{Perturbations}). Both of them are relevant for $V''_\star>0$ and solution becomes RG unstable. In addition these two modes become complex when $V''_\star>{9\over 4 L^2}$ and the solution is dynamically unstable.

\subsection{Electric/Magnetic solutions}

Including a magnetic or electric field the equations force the function C(r) (\ref{ra}) to be a fixed constant (see appendix \ref{appB}). The geometry is the same for all the following solutions, but the value $\phi_\star$ that the scalar assumes and the behavior of the charge density differ. The general solution for the metric has been found (in appendix \ref{appB}) and is studied in the next subsection \ref{geo}.

\subsubsection{Geometry of general solution}\label{geo}
The general solution for the metric is the following:
\be ds^2=L^2\left(-{f(r)\over r^2}dt^2+{dr^2\over f(r)r^2}\right)+C_0 dx_idx^i\ee
where
\be f(r)=1+K_1r+K_2r^2\sp C_0^2={1\over 2}L^2 E_\star\sp L^2={1\over V_\star}\sp E=h^2Z+{(q+hW)^2\over Z}\ee

The spatial coordinates of the field theory $x^i$ decouple from the radial and temporal coordinates. The coordinates $x^i$ enjoy rotational and translational symmetries, but not scaling symmetry.

The UV is located at $r\to 0$ where the geometry is asymptotically AdS$_2\times R^2$. Additionally, the metric is AdS$_2\times R^2$ for every value of $r$ if $K_1=K_2=0$. The latter case has been studied in \cite{gk}:
  \be ds^2=L^2\left(-dt^2+dr^2\over r^2\right)+C_0dx_idx^i\ee
Time scales the same way as the radial coordinate, however the spatial part does not scale.

The IR is located at $r\to \infty$ where the behavior of the metric depends on the two integration constants $K_1\sp K_2$. The scalar curvature is independent of the integration constants $K_1\sp K_2$ and equal to $-{2\over L^2}$. This means that, in principle, we can transform the metric to AdS$_2\times R^2$. However in some cases the transformation is singular and the metric contains a black hole. We consider different cases depending on the behavior of the function $f(r)$.
The function $f(r)$ is a quadratic polynomial with discriminant $\Delta=K_1^2-4K_2$. We distinguish the following cases regarding $\Delta$ and we focus on the AdS$_2$ component of the metric:

\begin{itemize}
\item $\Delta>0$: $f(r)$ has two real roots $ r_h^\pm={K_1\pm \sqrt{\Delta}\over 2}$. In this case we have an AdS$_2$ black hole with two horizons: the inner horizon is located at $r_h^-$ and the outer at $r_h^+$:
\be ds^2=L^2\left(-{K_2(r-r_h^-)(r-r_h^+)\over r^2}dt^2+{dr^2\over K_2(r-r_h^-)(r-r_h^+)r^2}\right)\ee

\item $\Delta=0$: This is the extremal limit of the previous case, where the two horizons coincide. We can get rid of the horizons with a simple redefinition of the radial coordinate:
\be ds^2={l^2\over u^2}(-d\tilde{t}^2+du^2)\ee
where $u={2r\over 2+K_1r}$. The metric is AdS$_2$ and has zero temperature.

\item $\Delta<0$: In this case $f(r)$ has no real roots. We can transform the metric to AdS$_2$ (see appendix \ref{appB}):

\be ds^2={\tilde{L}^2\over \cos^2 u}(-d\tau^2+du^2)\ee

%\{A metric of the form $G(\rho)(-dt^2+d\rho^2)$ has constant negative curvature $R=-{2\over L^2}$ for \be (G'^2 - G G'')=-{2\over L^2}G^3\eeThe general solution is\be G(r)={L^2c_1^2\over \cos^2{(\pm c_1(r+c_2))}}\ee}

\end{itemize}

%\begin{itemize}
%\item{$K_1<0$}: In this case $f(r)$ has one real root at $r_h=-{1/K_1}$. We have an AdS$_2$ black hole:

%\be ds^2=L^2\left(-{r-r_h\over r^2}dt^2+{dr^2\over (r-r_h)r^2}\right)\ee

%\item{$K_1>0$}: In this case $f(r)$ has no real roots and the geometry is simply AdS$_2$.
%\end{itemize}

% If $K_2\ne 0$ the geometry is asymptotically flat if we redefine $r=L\sqrt{K_2}\rho^{-1}\sp \tau=L\sqrt{K_2} t\sp x^i=\sqrt{C_0}y^i$.
%\be ds^2=-d\tau^2+{d\rho^2}+dy_idy^i \ee

 \subsubsection{Magnetic solution at zero charge density}\label{cs25}

 The simplest magnetic solution is obtained by setting $q=0\sp W=0$. In this case we have a magnetic field at zero charge density. The scalar field settles to a value $\phi_\star$ that, instead of the scalar potential $V$, extremizes the effective potential (\ref{Veff}) which reads:
 \be V'_{eff}(\phi_\star)=V'_\star-V_\star {Z'_\star\over Z_\star}=0\ee
 \subsubsection{Magnetic solution at finite charge density}\label{cs3}
 
Next, we study the magnetic solution at finite charge density which is obtained when $q=0\sp W\ne 0$.
 In this case the effective potential reads:
 \be V'_{eff}(\phi_\star)=V'_\star-V_\star {M'_\star\over M_\star}=0\ee
 Where \be M_\star=Z_\star+{W_\star^2\over Z_\star}\ee
  The value of the scalar, $\phi_\star$ does not depend on the value of the magnetic field, $h$.
 
 When $W$ is non-zero there is a finite IR charge density:
  \be A_t=\m-{Q\over r}\sp \m={\rm constant}\sp Q^{2}={2W_\star^2\over V_\star Z_\star(W_\star^2+Z_\star^2)}\ee
   The charge density, Q, is fixed by the equations and is independent of the magnetic field. It depends only on the values of the coupling functions at $\phi_\star$. When $W_\star=0$ the charge density is zero.

\subsubsection{Electric solution at finite charge density}\label{cs2}

In this case we only have an electric field ($h=0$). This case is the electric dual of the solution \ref{cs25} with $W=0$ and has been studied in \cite{gk}. The scalar field settles to a value $\phi_\star$ that extremizes the effective potential:
 \be V_{eff}'(\phi_\star)=V'_\star+V_\star {Z'_\star\over Z_\star}\ee
The solution is again AdS$_2\times R^2$ with finite charge density:
 \be A_t=\m-{Q\over r}\sp \m={\rm constant}\sp Q^2={2\over V_\star Z_\star} \ee
 The IR charge density is again fixed by the equations and is independent of the electric flux. It depends only on the values of the coupling functions at $\phi_\star$. In this purely electric solution there is always a non-zero charge density, since $Q$ cannot be zero.

\subsubsection{Dyonic solution at zero/finite charge density}\label{cs4}

We now consider the most general solution with finite electric and magnetic fluxes ($q\ne 0\sp h\ne 0$). This is the electromagnetic dual of case \ref{cs25} with $W\ne 0$. In this case, the scalar field settles to a value $\phi_\star$ that extremizes the effective potential:

\be V'_{eff}(\phi_\star)=V'_\star-{E'_\star\over E_\star}V_\star\ee
 Where \be E_\star=h^2Z_\star+{(q+hW_\star)^2\over Z_\star}\ee
 The value $\phi_\star$ that the scalar assumes now depends on the values of the electric ($q$) and magnetic ($h$) fluxes.

The charge density is given by:
\be A_t=\m-{Q\over r}\sp \m={\rm constant}\sp {1\over Q^2}={V_\star Z_\star\over 2}\left(1+\left({hZ_\star\over q+hW_\star}\right)^2\right)\ee

The charge density is fixed. In contrast to all the previous cases it now depends on the value of the magnetic field, $h$.

Consider now the special case $q+hW_\star=0$. The effective potential is:
\be V'_{eff}(\phi_\star)=V'_\star-{Z'_\star\over Z_\star}V_\star\ee
and is independent of $q,h$. The charge density, $Q$, is now zero. This means that when the magnetic field assumes the specific value $h=-{q\over W_\star}$ in terms of the electric flux, the charge density, $Q$, vanishes. This is similar to what is expected to happen in the quantum Hall effect.
\newpage
\section{Running Scalar Solutions\label{SS}}
We now assume that the scalar field runs logarithmically: $$\phi=\phi_0+a \log(r)$$ to $\pm\infty$ in the IR and that asymptotically the coupling functions behave as:
\be Z(\phi)=Z_0 e^{\gamma\phi}, V(\phi)=V_0 e^{-\delta\phi}, W(\phi)=e^{\chi\phi}\ee

We first consider the pure EMD theory discarding the PQ term $F\wedge F$ of the action. This theory has already been studied in the past, \cite{gk}, \cite{eht},\cite{EMD}. Our ansatz is different, as it also contains the magnetic field. The solutions come in pairs, due to the electromagnetic duality of the EMD theory.

We next consider the case with intrinsic parity violation, where we keep all the terms of the action $(W\ne 0)$. In general we expect new solutions.

We look for metrics with a dynamical exponent $z$ and a hyperscaling violating exponent $\theta$:
\be ds^2=-r^{\theta-2z}dt^2+B_0 r^{\theta-2}dr^2+r^{\theta-2}(dx_1^2+dx_2^2)\ee
while for the gauge field we assume the scaling form
\be A_t=\m+A_0r^{\zeta-z}\ee
The triplet of critical exponents $(\theta,z,\zeta)$ determines the geometry and the behavior of the charge density and is, in turn, determined by the exponents $\g,\d,\chi$ that appear in the asymptotic behavior of the coupling functions. The constants $\m, Q$ correspond to the chemical potential and charge density of the dual QFT respectively.

The solutions with $V$ subleading in the IR were also found. Such solutions seem problematic because not only are there important parameters of the solution that cannot be fixed, but also the IR is never well-defined for any values of the exponents $\g,\d$. It is not known if they are able to describe real physical systems, however we present them in appendix \ref{V0} for completeness.

\subsection{Scaling solutions without intrinsic P-violation}\label{W0}

In this section we present the running scalar solutions of the EMD theory ignoring the PQ term ($W=0$). The charge density in this case is closely related to the electric flux. Solutions with an electric flux always have a finite charge density, while solutions without electric flux always have zero charge density. We find a neutral and an electric solution which have been recently studied in \cite{gk}. We also find the magnetic dual of the electric solution. They are connected by $\g\to -\g\sp q^2/Z_0\leftrightarrow h^2Z_0$. In the special case $\g=0$ we have a dyonic solution which is electromagnetically self-dual.

\subsubsection{Neutral solution}\label{Ic}

In this case the gauge field is zero and the Lorentz invariance of the field theory is restored ($z=1$). This solution has been studied in the past, \cite{gk},\cite{eht},\cite{gk2},\cite{EMD}.
\be \begin{split}
& ds^2=r^\theta\left({-dt^2+B_0dr^2+dx_idx^i\over r^2}\right)\sp e^\phi=e^{\phi_0}r^{\sqrt{\theta(\theta-2)}}\\
& \theta={2\d^2\over \d^2-1}\sp z=1\sp B_0={2e^{\d\phi_0}\over V_0}{3-\delta^2\over (\d^2-1)^2}\\
\end{split}\ee
The thermal entropy of the field theory for small temperatures scales as:
\be S\sim T^{2-\theta\over z}\sim T^{2\over 1- \d^2}\ee
When the exponent is negative the field theory has a mass gap and a discrete spectrum, \cite{eht}.

Turning on the t-component of the gauge field in this background we find
\be A_t=\m+A_1 r^{1-\g\sqrt{\theta(\theta-2)}}\ee
with $A_1\sim q$.
The second term creates a constant finite electric flux in the IR when $q\ne 0$.

The singularity is always located in the IR. However the location of the IR (as well as the behavior of the entropy) depends on the value of $\d$:
\begin{itemize}
\item{$\d^2<1$}:
The IR is located at $r\to\infty$. The entropy scales with a positive power of T and vanishes at zero temperature. In the special case $\d=0$ we arrive at the AdS$_4$ solution with constant scalar (in section \ref{CS1}). According to \cite{eht} the spectrum of the theory is continuous without a mass gap.

\item{$\d^2=1$}:
The geometry is AdS$_4$ with constant scalar (same as section \ref{CS1}). The thermal entropy scales as $\sim T^2$.

\item{$1<\d^2<3$}:
The IR is located at $r\to 0$. The thermal entropy scales with a negative power of temperature and becomes very large for zero temperature. The boundary theory has a discrete spectrum with a mass gap, \cite{EMD}, \cite{eht}.

\item{$\d^2=3$}:
This value can only be reached by allowing $V$ to be subleading. The solution is presented in appendix \ref{V0}.

\item{$\d^2>3$}:
Such values are unacceptable since the r-coordinate becomes time-like. They violate the Gubser bound \cite{Singularities}.

\end{itemize}

\subsubsection{Magnetic solution at zero charge density\label{IIa}}

\begin{figure}[H]
\centering
\includegraphics[width=0.6\textwidth]{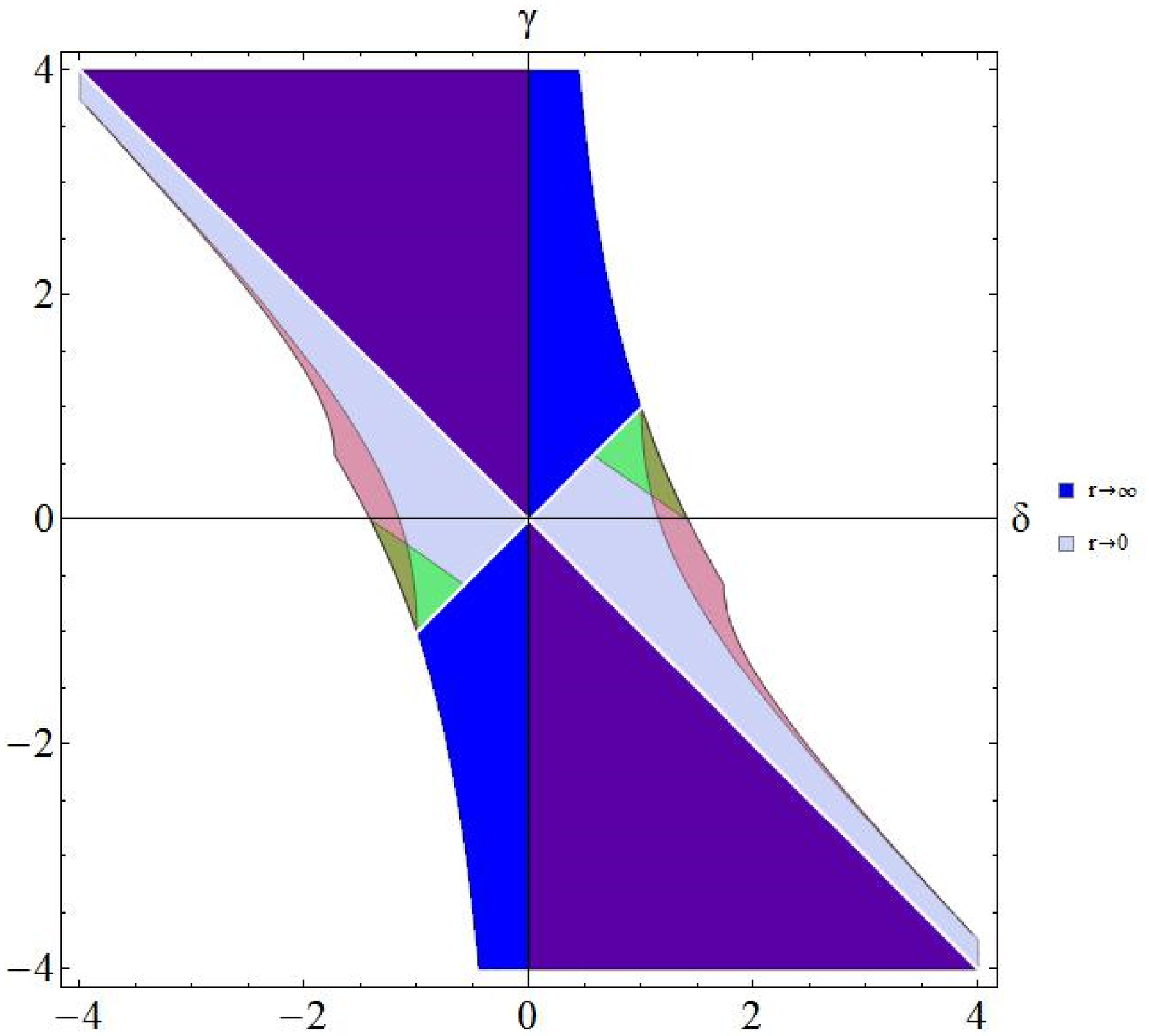}
\caption{The region of validity of the magnetic solution \protect\ref{IIa} is shown above. Within the light blue area $r\xrightarrow[IR]{}0$ while in the dark blue area $r\xrightarrow[IR]{}\infty$. The solution is valid for $r\to\infty$ only if the electric flux is zero to all orders (see appendix \protect\ref{appC}). The diagonals $\g=-\d,\g=\d$ correspond to AdS$_2\times R^2$ and conformally AdS$_2\times R^2$ geometries respectively. The singularity is in most cases located in the IR, except for the values within the purple area. In the light red area the entropy scales with a negative power of temperature, which indicates that the field theory may have a mass gap. $A_t'$ is relevant within the green area and irrelevant everywhere else.}
\label{magnetic}
\end{figure}

In this case the magnetic field is finite and the electric flux is zero. The Lifshitz exponent now depends on the parameters $\g,\d$.

\be\begin{split} & ds^2=-r^{\theta-2z}dt^2+r^{\theta-2}\left(B_0dr^2+dx_idx^i\right)\sp  e^{\phi}=e^{\phi_0}r^{\theta/\d}\\
& \theta={4\d\over \delta-\gamma}\sp z={3\d^2-\g^2+2\g\d-4\over \d^2-\g^2}\\
& B_0=(2+z-\theta) \left({1+z-\theta\over V_0}\right)^{\g\over \g+\d}\left({2(z-1)\over h^2 Z_0}\right)^{\d\over \g+\d}\sp e^{(\g+\d)\phi_0}=2{z-1\over 1+z-\theta}{V_0\over h^2Z_0}\end{split}\ee
% B_0={(\theta - z - 2) (\theta - z - 1)\over V_0}e^{\d\phi_0}\sp h^2=2{z-1\over 1+z-\theta}{V_0\over Z_0}e^{-(\g+\d)\phi_0}
%
The thermal entropy of the field theory scales at low temperatures as
\be S\sim T^{2-\theta\over z}\sim T^{2(\g+\d)^2\over \g^2-2\g\d-3\d^2+4}\ee
According to \cite{eht} a negative exponent indicates an unstable black hole, as the entropy becomes very large at extremality. It also indicates that the field theory may have a discrete spectrum with a mass gap, \cite{QCD8}. There is a finite entropy at extremality in the special case $\g=-\d$, which corresponds to an AdS$_2\times R^2$ space-time according to \cite{gk}.
The parameter values for which the exponent is negative are plotted in figure \ref{magnetic}.

When the magnetic flux becomes zero ($z=1\Rightarrow \g\d={2-\d^2}$) we arrive at the previous case \ref{Ic}. In the special case $\d=0$ the metric becomes Lifshitz and the exponent of the entropy is positive for any $\g$. We can also arrive at a conformally Rindler metric when $z=0$. This corresponds to the curves $(\g-\d)^2=4(\d^2-1)$ on the $\d-\g$ plane. These curves separate the thermodynamically stable and unstable regions on the $\d-\g$ plane. The behavior of the entropy in these cases is not known and needs to be studied further. What can be done in this case is discussed in \cite{QCD8}, \cite{Gur}.

Turning on the t-component of the gauge field in this background we obtain:
\be A_t=\m+A_1 r^{2-z-\theta{\g\over\d}}\ee
The amplitude is proportional to $A_1\sim q$. The region of the parameter space where $A_t'$ is relevant is shown in figure \ref{magnetic}.

We now consider the cases $\d=\pm \g$, \cite{gk}, which make the exponents $\theta,z$ blow up:
\begin{itemize}

\item{$\g=-\d$}: In this case $z\to\infty$ while $\theta$ remains finite. Changing the radial coordinate $r^z=\rho$ we obtain the AdS$_2\times R^2$ geometry with constant scalar.

\item{$\g=\d$}: In this case $z\to\infty\sp \theta\to \infty$, while their ratio ${\theta\over z}=\l$ is finite. After changing the radial coordinate $r^z=\rho$ the metric becomes:
\be ds^2=\rho^{\l}({d\rho^2-dt^2\over \rho^2}+dx_idx^i)\ee
which is conformally AdS$_2\times R^2$. This solution has some interesting properties and for this it has been named semi-locally critical \cite{gk}.

\end{itemize}

\subsubsection{Electric solution at finite charge density\label{Ia}}

In this case the electric flux is finite and the magnetic field is zero. It has been studied in the past in \cite{gk},\cite{eht},\cite{EMD}. This is the electric dual of case \ref{IIa} with $W=0$, which means that the geometry and scalar of this solution can be obtained with the duality transformation $\g\to-\g\sp  h^2 Z_0\to q^2/Z_0$. The analysis of this solution is analogous to \ref{IIa}. The only difference is that there is no magnetic field and the charge density is now non-zero.

\be\begin{split} & A_t=\m+q^{\g\over \d-\g}\sqrt{(2z-2)^{2\g-\d\over \g-\d}(1+z-\theta)^{\g\over \d-\g}Z_0^{\d\over \g-\d}\over V_0^{\g\over \g-\d}(2+z-\theta)}r^{\theta-2-z}\\
& \theta={4\d\over \delta+\gamma}\sp z={3\d^2-\g^2-2\g\d-4\over \d^2-\g^2}\end{split}\ee
%$ A_t=\m+\sqrt{2(z-1)\over Z_0 e^{\g\phi_0}(2+z-\theta)}r^{\theta-2-z}$
%$A_0=(q+hW_0)^{\g\over \d-\g}\sqrt{(2z-2)^{2\g-\d\over \g-\d}(1+z-\theta)^{\g\over \d-\g}Z_0^{\d\over \g-\d}\over V_0^{\g\over \g-\d}(2+z-\theta)}$
% A_t=\m+q^{\g\over \g-\d}\sqrt{2(z-1)^{\d\over \d-\g}(1+z-\theta)^{\g\over \g-\d}\over Z_0^{2\g-\d\over \g-\d}V_0^{\g\over \g-\d}(2+z-\theta)}r^{\theta-2-z}

\begin{figure}[H]
\centering
\includegraphics[width=0.6\textwidth]{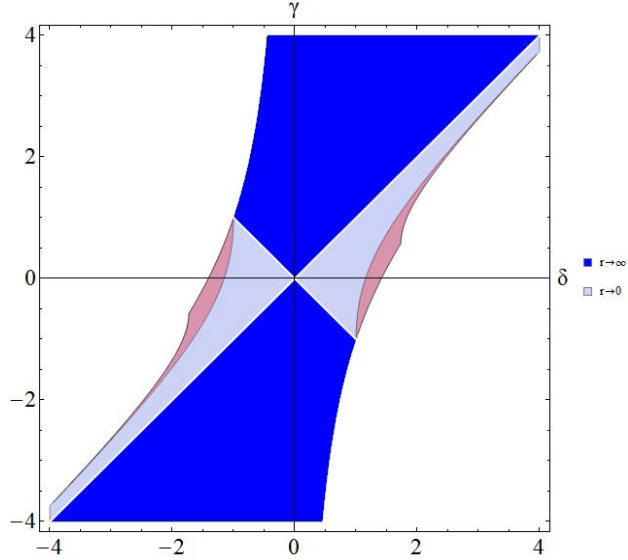}
\caption{The region of validity of the electric solution \protect\ref{Ia} is shown above. Within the light blue area $r\xrightarrow[IR]{}0$ while in the dark blue $r\xrightarrow[IR]{}\infty$. The solution is valid in the dark blue area only if the magnetic field is zero to all orders (see appendix \protect\ref{appC}). In the light red area the IR is still at $r\xrightarrow[IR]{}0$, but the entropy scales with a negative power of temperature, which means that the boundary theory has a mass gap.}
\label{electric}
\end{figure}
The charge density assumes a value proportional to a power of the electric flux:
\be A_0\sim q^{\g\over \d-\g}\sim q^{4-\theta\over 2(\theta-2)}\ee
The entropy of the field theory scales at low temperatures as
\be S\sim T^{2-\theta\over z}\sim T^{2(\g-\d)^2\over \g^2+2\g\d-3\d^2+4}\ee
The sign of the exponent is plotted in figure \ref{electric}.

\subsubsection{Dyonic solution at finite charge density}\label{IIIa}

In the leading order solution both electric and magnetic fluxes are finite if $Z$ is constant, which means that $\g=0$, if we require the scalar field to run to infinity. This solution is electromagnetically self-dual.
\be\begin{split} & ds^2=-r^{4-2z}dt^2+r^{2}\left(B_0dr^2+dx_idx^i\right)\sp  e^{\phi}=e^{\phi_0}r^{4/\delta}\sp A_t=\m+\sqrt{2(z-1)\over Z_0(2+z-\theta)}r^{\zeta-z}\\
&  \theta=4\sp z={3\d^2-4\over \d^2}\sp \zeta=2\\
& B_0={(2+z-\theta)(z-1)}{2Z_0\over q^2+h^2Z_0^2}\sp e^{\d\phi_0}=2{z-1\over 1+z-\theta}{V_0Z_0\over q^2+h^2Z_0^2}\end{split}\ee
%B_0={(\theta - z - 2) (\theta - z - 1)\over V_0}e^{\d\phi_0}\sp q^2+h^2Z_0^2=2{z-1\over 1+z-\theta}{V_0Z_0}e^{-\d\phi_0}
%B_0=(2+z-\theta) \left({1+z-\theta\over V_0}\right)^{\g\over \g+\d}\left({z-1\over h^2 Z_0}\right)^{\d\over \g+\d}\sp e^{(\g+\d)\phi_0}=2{z-1\over 1+z-\theta}{V_0\over h^2Z_0}

The charge density in this case is not affected by the existence of the magnetic field. However when the charge density is zero ($z=1$) both the electric flux and the magnetic field must be zero.

The IR is located at $r\to 0$ if $\d^2<2$.
Values $2<\d^2<4$ are forbidden by the null energy condition (appendix \ref{NEC}), while values $\d^2>4$ are forbidden by the Gubser bound. The singularity is always located at $r\to 0$.

The entropy of the field theory scales at low temperatures as
\be S\sim T^{2-\theta\over z}\sim T^{2\d^2\over 4-3\d^2}\ee
For $\d^2<{4\over 3}$ the entropy scales with a positive power of temperature. When ${4\over 3}<\d^2<2$ the exponent is negative and the dual field theory may be gapped, \cite{QCD8}.

Now consider the case $\d=0$ in which $z\to\infty$. Redefining the radial coordinate as: $r^z=\rho$ we obtain an AdS$_2\times R^2$ space-time with constant scalar:
\be ds^2={1\over \rho^2}\left(-dt^2+B_\star d\rho^2\right)+dx_idx^i\sp B_\star=\lim\limits_{z\to\infty}{B_0\over z^2}={1\over V_0}\ee

\newpage
\subsection{Scaling solutions with intrinsic P-violation}
In this section we study solutions with $W\ne 0$. When $W$ is subleading we obtain the same leading order solutions studied in the previous section. We correct these cases to first order of $W$ by including a subleading term in appendix \ref{PQp}. In the present section we study the solutions when $W$ participates in the equations to leading order.

\subsubsection{Dyonic solution at zero/finite charge density}
There is a dyonic solution when $W$ is constant and leading in the IR (this implies that $\chi=0$). This is the electromagnetic dual of solution \ref{IIa} with $W\ne 0$.
\be\begin{split} & ds^2=-r^{\theta-2z}dt^2+r^{\theta-2}\left(B_0dr^2+dx_idx^i\right)\sp  e^{\phi}=e^{\phi_0}r^{\theta/\d}\sp A_t=\m+A_0r^{\zeta-z}\\
& \theta={4\d\over \delta+\gamma}\sp z={3\d^2-\g^2-2\g\d-4\over \d^2-\g^2}\sp \zeta=\theta-2\\
& B_0={(2+z-\theta ) \left({1+z-\theta \over V_0}\right)^{\g\over \g-\d}}\left({ 2(z-1)Z_0\over (q+hW_0)^2}\right)^{\d\over \d-\g}\sp e^{(\d-\g)\phi_0}=2{z-1\over 1+z-\theta}{V_0 Z_0\over (q+hW_0)^2}\\
& A_0=(q+hW_0)^{\g\over \d-\g}\sqrt{(2z-2)^{2\g-\d\over \g-\d}(1+z-\theta)^{\g\over \d-\g}Z_0^{\d\over \g-\d}\over V_0^{\g\over \g-\d}(2+z-\theta)} \end{split}\ee
The values of the scalar and the charge density both depend on the electric and magnetic fluxes. The values of the exponents $\theta,z,\zeta$ are the same as in case \ref{Ia}. The analysis of this solution is identical to case \ref{Ia}, however this solution shows an interesting behavior when $z=1$ which we will now consider. Setting $z=1$ (which is equivalent to setting $\g=\d-2/\d$) in the above equations we obtain:

\be \begin{split}
& ds^2=r^\theta\left({-dt^2+B_0dr^2+dx_idx^i\over r^2}\right)\sp e^\phi=e^{\phi_0}r^{\theta/\d}\\
& \theta={2\d^2\over \d^2-1}\sp z=1\sp B_0={2e^{\d\phi_0}\over V_0}{3-\delta^2\over (\d^2-1)^2}\sp q+hW_0=0\\
\end{split}\ee

This is the generalization of the neutral solution studied in section \ref{Ic} when $W\ne 0$. The important difference is that instead of $q=h=0$ we have the less strict requirement $q+hW_0=0$. This is satisfied, of course, when both $q,h$ are zero, but it is also satisfied for non-zero values of $q,h$. The important conclusion is that when the magnetic field assumes a specific value in terms of the electric flux $h=-{q\over W_0}$ the charge density vanishes. This behavior also appears in the constant scalar case \ref{cs4} on an AdS$_2\times R^2$ background.
\newpage
\subsubsection{Magnetic solution at finite charge density\label{IIa2}}

In this case $W$ participates in the leading order solution and $\chi$ is, in general, non-zero. This is a stable magnetic quantum critical line, characterized by 3 exponents $(\g,\d,\chi)$. In previous cases the conductivity exponent, $\zeta$ was dependent on the hyperscaling violating exponent $\theta$. In this case the 3 exponents $(z,\theta,\zeta)$ are completely independent from each other.
\be\begin{split}& ds^2=-r^{\theta-2z}dt^2+r^{\theta-2}\left(B_0dr^2+dx_idx^i\right) \sp  e^{\phi}=e^{\phi_0}r^{\theta\over\delta}\sp A_t=\m+A_0 r^{\zeta-z}\\
& \theta={4\d\over \g+\d-2\chi}\sp z={3\d^2-(\g-2\chi)^2-2\d(\g-2\chi)-4\over \d^2-(\g-2\chi)^2}\sp \zeta=2{\d-\g\over \g+\d-2\chi}\\
& e^{(\d+2\chi-\g)\phi_0}=2{z-1\over 1+z-\theta}{Z_0V_0\over h^2W_0^2}\sp B_0=\left({1+z-\theta\over V_0}\right)^{2\chi-\g\over \d+2\chi-\g}(2+z-\theta)\left({2(z-1)Z_0\over h^2 W_0 ^2}\right)^{\d\over \d+2\chi-\g}\\
&  A_0={(hW_0)^{\g\over \d+2\chi-\g}\over \zeta-z}\sqrt{(2z-2)^{2\g-\d-2\chi\over \g-\d-2\chi}(2+z-\theta)\left({V_0\over 1+z-\theta}\right)^{\g\over \g-\d-2\chi}Z_0^{\d+2\chi\over \g-\d-2\chi} }\end{split}\ee

%$h^2=2{z-1\over 1+z-\theta}{Z_0V_0\over W_0^2}e^{(\g-\d-2\chi)\phi_0}\sp A_0={1\over \zeta-z}\sqrt{2(1-z)(\theta-z-2)\over Z_0 e^{\g\phi_0}}$

The entropy at low temperatures scales as:
\be S\sim T^{2-\theta\over z}\sim T^{2 (2 \chi - \g + \d)^2\over 4 + 4 \chi^2 + \g^2 + 2 \g \d - 3 \d^2 - 4 \chi (\g + \d)}\label{ent}\ee
For values of $\chi$: $ {\g + \d\over 2} -  \sqrt{ \d^2-1}<\chi<{\g + \d\over 2} +  \sqrt{ \d^2-1}$ the entropy scales with a negative power of temperature and the boundary theory may have a mass gap, \cite{QCD8},\cite{eht}.
The charge density is fixed and proportional to a power of the magnetic field
\be A_0 \sim h^{\g\over \d+2\chi-\g}\sim h^{\theta-2\zeta\over 2(\theta-2)}\ee
which shows a similarity to the results in \cite{magn}.

The parameter $\chi$ must obey a constraint for this solution to be valid in the IR. This constraint comes from the requirement that $W/Z\to \infty$ in the IR (for details see appendix \ref{appC}) and is the following:
\bsea (\g+\d-2\chi)(\chi-\g)>0\sp r\xrightarrow[IR]{}\infty \\
(\g+\d-2\chi)(\chi-\g)<0\sp r\xrightarrow[IR]{}0
\label{WZ}\esea

The behavior of $W\sim r^{a\chi}$ in the IR depends on the value of $\chi$ relative to the other two exponents $\g,\d$.
\begin{itemize}
\item{$\chi(\chi-{\g+\d\over 2})<0$}: Then $W\to \infty$ if $r\xrightarrow[IR]{}\infty$ and $W\to 0$ if $r\xrightarrow[IR]{}0$.
\item{$\chi(\chi-{\g+\d\over 2})>0$}:  Then $W\to 0$ if $r\xrightarrow[IR]{}\infty$ and $W\to \infty$ if $r\xrightarrow[IR]{}0$.
\end{itemize}
When the integration constant $q$ (\ref{q}) is non-zero we must require $W\to\infty$ in the IR and we have an additional constraint. In this case $q$ does not appear in the leading order solution.
When $q=0$ this constraint is absent.

When $\chi$ assumes special values we arrive at some interesting cases:
\begin{itemize}
\item{$\chi=\g\sp W_0=Z_0$}: We obtain the magnetic solution of section \ref{IIa}.
\item{$\chi=0\sp hW_0\to q$}: In this case we obtain the electric solution of section \ref{Ia}.
\item{$\chi={2+\g\d-\d^2\over 2\d}$} In this case ($h=0=A_0\to z=1$) we obtain the neutral solution of section \ref{Ic}.
\item{$\chi\to \infty$}: In this case we arrive at the neutral AdS$_4$ solution with constant scalar of section \ref{CS1}.
\item{$\chi={\g-\d\over 2}$}: In this case $z\to\infty$ while $\theta$ is finite. Redefining the radial coordinate $r^z=\rho$ the geometry becomes AdS$_2\times R^2$ with constant scalar. Also the entropy (\ref{ent}) is constant and independent of the temperature. We have a finite entropy at extremality.
\item{$\chi={\g+\d\over 2}$}: In this case $\theta,z\to \infty$, while their ratio is constant. Redefining the radial coordinate $r^z=\rho$ the geometry becomes conformally AdS$_2\times R^2$.
\be ds^2=\rho^{\l}({d\rho^2-dt^2\over \rho^2}+dx_idx^i)\sp \l={\theta\over z}={2\d^2\over \d^2-1}\ee
 This is the semi-locally critical solution which appeared in \cite{gk} in the absence of magnetic fields.
\item{$\chi={\g + \d\over 2} \pm  \sqrt{-1 + \d^2}$}: In these cases $z=0$ and the space-time is conformally Rindler. The curves corresponding to this case separate the thermodynamically stable and unstable regions in the $\d-\g$ plane. The formula for the entropy (\ref{ent}) no longer holds in this case and further studies are required, \cite{QCD8}, \cite{Gur}.
\end{itemize}

Studying the linear perturbations around this solution we find two pairs of modes summing to $2+z-\theta$. The first pair corresponds to the finite temperature perturbation and the marginal perturbation, the latter of which can be absorbed by rescaling the coordinates. The other two modes $b_\pm$ are non-universal and also sum to $2+z-\theta$. Their forms are not very enlightening and are presented in section \ref{P1} of appendix \ref{Perturbations}. One of these modes is always positive, while the other is always negative. The solution is, therefore, RG stable. The values of $z,\theta$ which make $b_\pm$ complex are forbidden by combining the null-energy condition \ref{nec} and the conditions for a well-defined IR (see appendix \ref{HVM}), which shows that this solution is also dynamically stable.

\begin{figure}[H]
\centering
\includegraphics[width=0.49\textwidth]{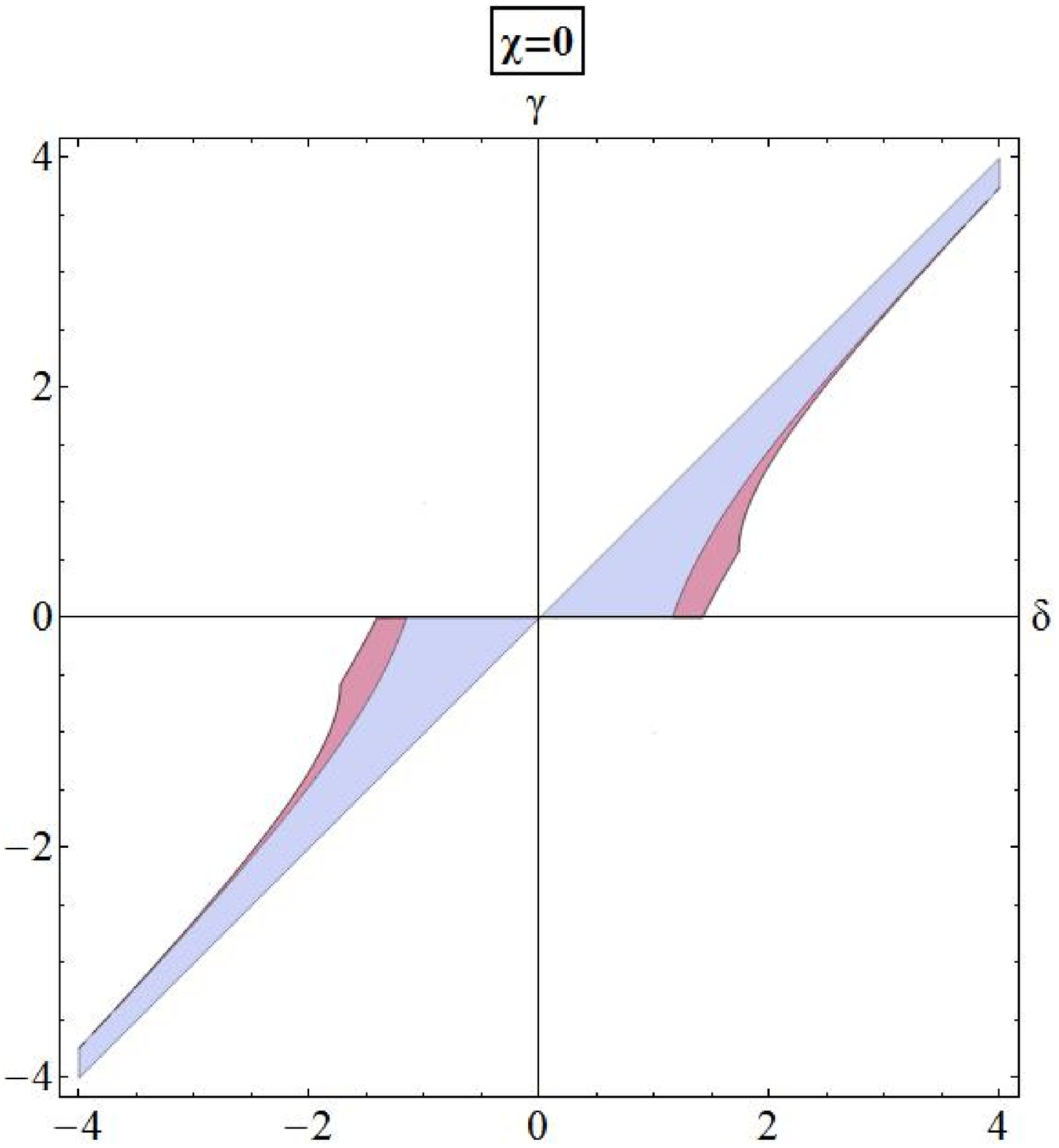}
\includegraphics[width=0.49\textwidth]{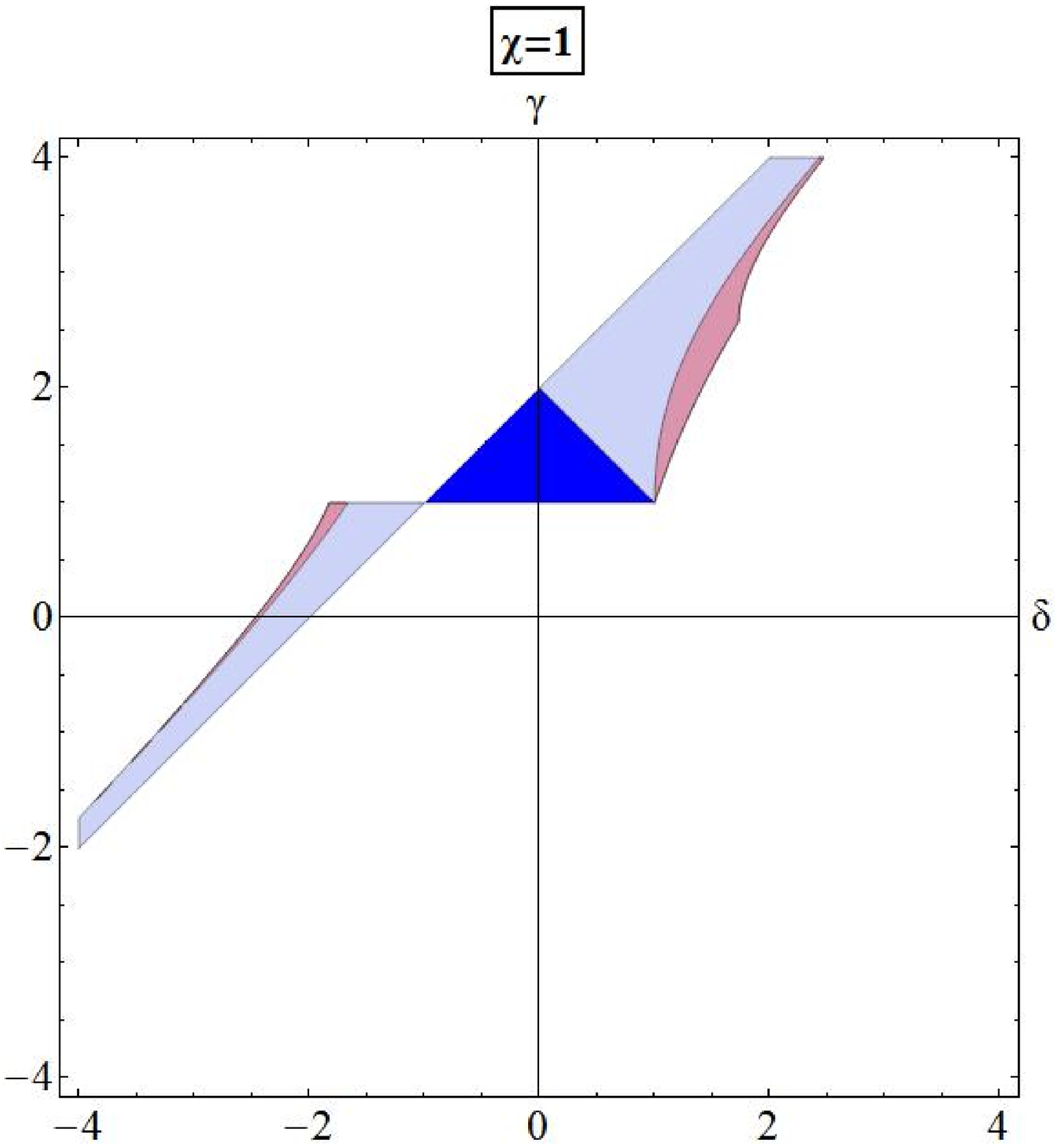}
\includegraphics[width=0.49\textwidth]{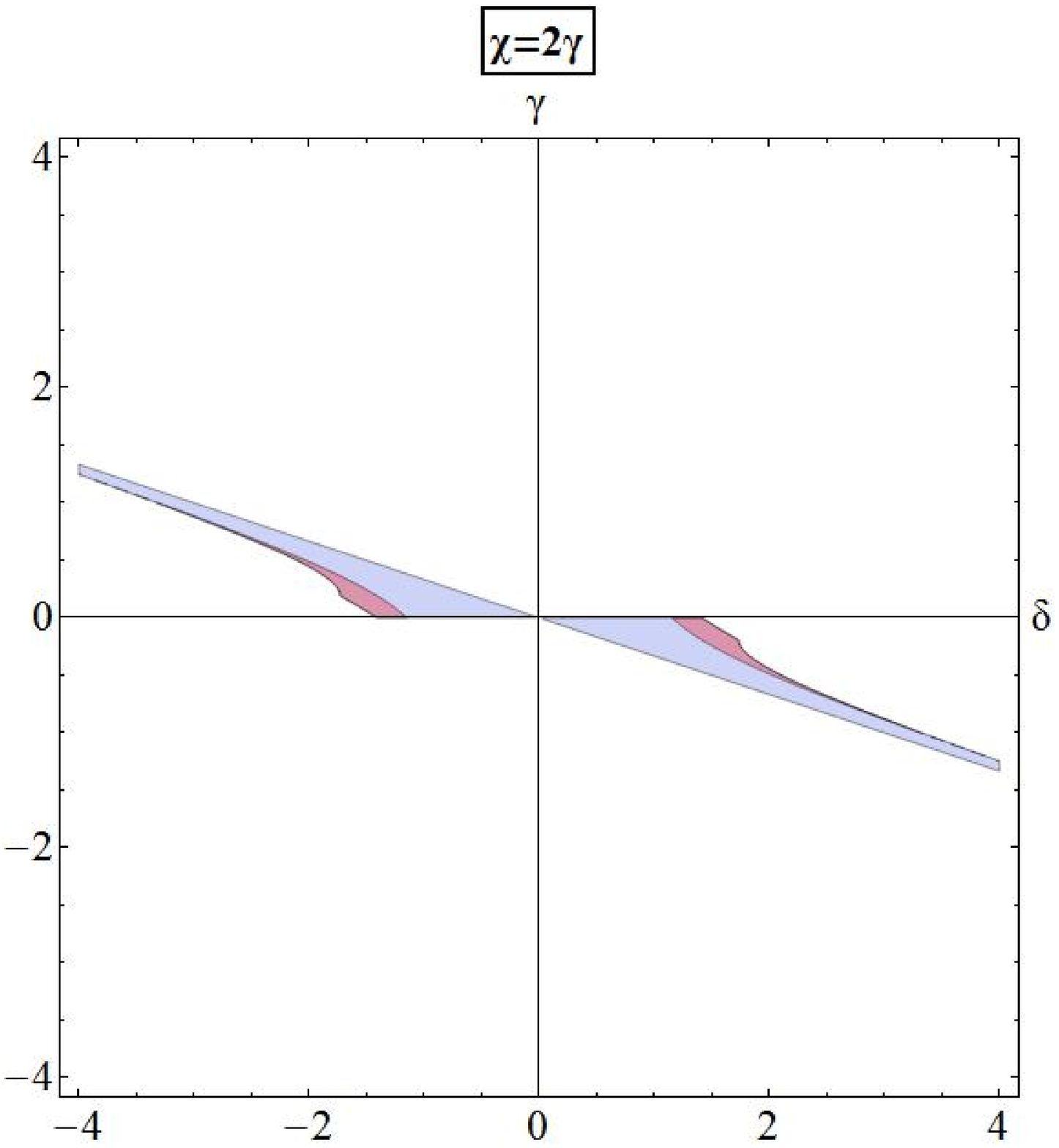}
\includegraphics[width=0.49\textwidth]{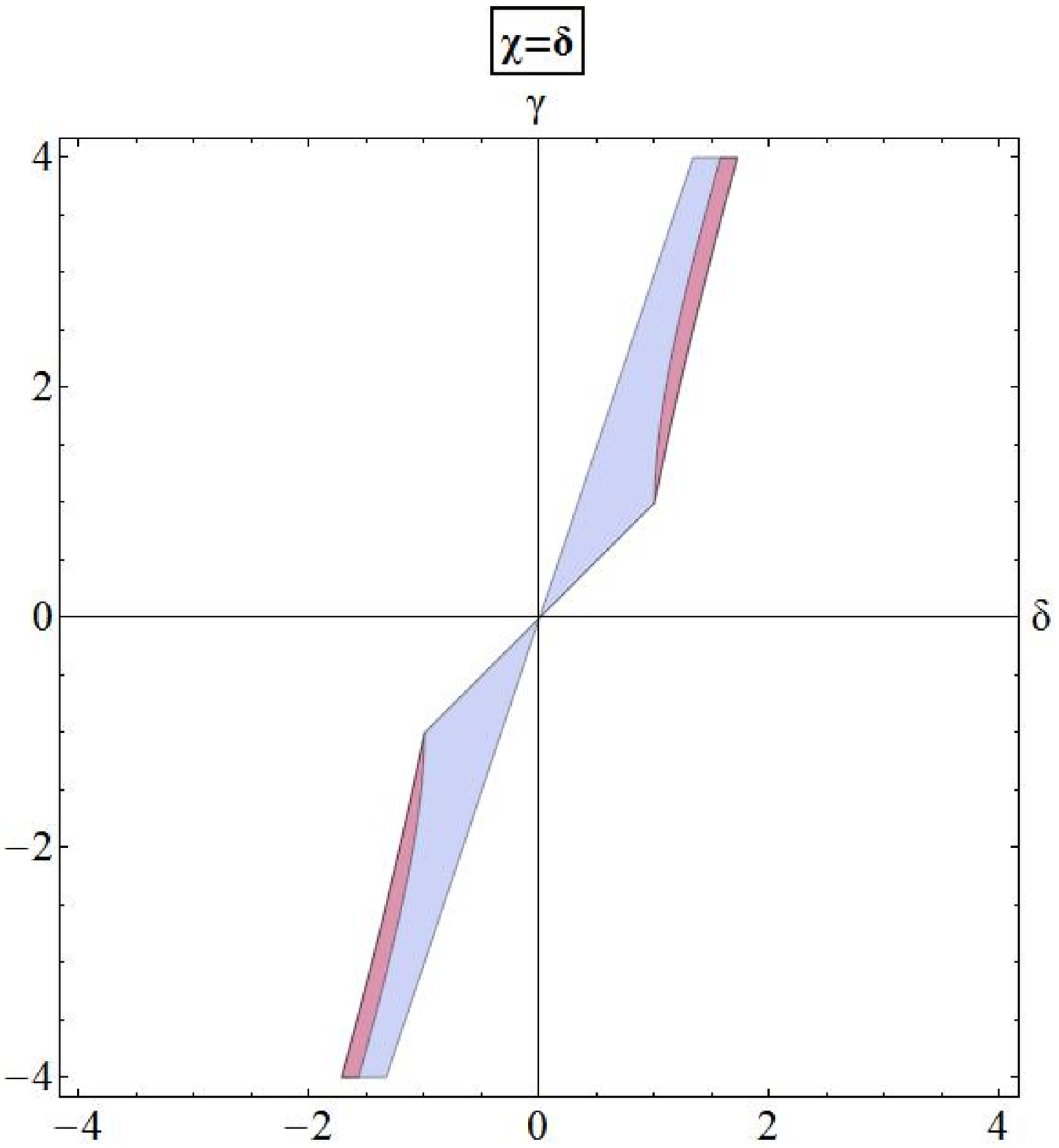}
\caption{\small The region of validity of the solution \protect\ref{IIa2} is shown above in the $\d- \g$ plane for various values of $\chi$.  Within the light blue area the IR is located at $r\xrightarrow[IR]{}0$ while in the dark blue it is located at $r\xrightarrow[IR]{}\infty$. For values within the light red area the field theory is expected to have a finite mass gap (as the entropy scales with a negative power of temperature, \cite{QCD8}).}
\label{f2}
\end{figure}

Finally, in figure \ref{f2} we present the region of the $\g-\d$ plane where this solution is valid as well as the location of the IR for various values of $\chi$. The constraints used are the following:
\begin{enumerate}[I]
\item{\bf Null-energy condition}: The null-energy condition \ref{NEC} is satisfied for $(2 - 2 z + \theta) (\theta-2)\geq 0 \sp (z-1) (2 + z - \theta)\geq 0$.
\item{\bf Well-defined IR}: The IR is well-defined (see appendix \ref{HVM}) when $(\theta-2)(\theta-2z)>0$ and is located at $r\to 0$ if $2+z-\theta<0$ or at $r\to\infty$ if $2+z-\theta>0$.
\item{\bf $W/Z\to\infty$}: This combination must be leading in the IR in order to arrive at this solution. The constraint is given by (\ref{WZ}).
\item{\bf Space-like radial coordinate (Gubser bound)}: $B_0$ must be positive which means $(\theta-z-2)(\theta-z-1)>0$.
\item{\bf Real magnetic field}: We require $h^2>0$ which translates to $(z-1)(1+z-\theta)>0$.
\item{\bf Real charge density $A_0$}: This requirement gives $(z-1)(2+z-\theta)>0$ and is already satisfied if the null-energy condition holds.

\end{enumerate}

\section{Conclusion}

In the constant scalar case in the absence of magnetic fields we confirmed previous studies, \cite{gk}, which include a neutral phase on AdS$_4$ and a charged phase on AdS$_2\times R^2$.

Turning on the magnetic field we obtain an AdS$_2\times R^2$ geometry both at zero and finite charge density, which may contain a black hole with a pair of horizons. In both of these cases there are RG instabilities originating in the scalar and gauge fields. We also find a dyonic solution in which the charge density depends on the magnetic field. In this case the charge density vanishes for a special value of the magnetic field.

In the running scalar case, in the absence of the PQ term ($W=0$), we found the usual neutral and electric solutions studied in the past \cite{gk},\cite{eht},\cite{gk2},\cite{EMD} along with their magnetic duals. They are connected by the transformation $\g\to-\g\sp q^2/Z_0\leftrightarrow h^2Z_0$. We also found a dyonic solution which is electromagnetically self-dual. In this case the magnetic field does not affect the value of the charge density.

When $W$ is constant and leading, we found a stable dyonic critical line in which the value of the charge density is affected by the presence of the magnetic field. The charge density vanishes when the magnetic field assumes a specific value in terms of the electric flux.

 We also found a stable magnetic critical line at finite density characterized by three exponents ($\g,\d,\chi$). In this case the charge density is proportional to a power of the magnetic field, which depends on the values of two critical exponents $\theta,\zeta$.

\addcontentsline{toc}{section}{Acknowledgements}
\section*{Acknowledgements}

I would like to thank Elias Kiritsis for suggesting the topic and for helpful conversations, comments and suggestions during the course of this work. I would also like to acknowledge helpful conversations with Chris Rosen.

\newpage
\appendix
\renewcommand{\theequation}{\thesection.\arabic{equation}}
\addcontentsline{toc}{section}{Appendix}
\section*{Appendix}

\section{The anti-de Sitter space-time \label{AdS}}

The anti-de Sitter space is the maximally symmetric solution of the Einstein equations with a negative cosmological constant. Consider the Einstein-Hilbert action with a cosmological constant $\L$ in d+1 dimensions:
\be S={1\over 16\pi G_N} \int \ud^{d+1}x \sqrt{-\det g}(R-2\L)\ee
By varying this action we obtain the following Einstein equations:
\be R_{\m\n}-{1\over 2} g_{\m\n} R=-\L g_{\m\n}\ee
Taking the trace we obtain the scalar curvature:
\be R=2 {d+1\over d-1}\L\ee
The scalar curvature is constant and its sign depends on the sign of the cosmological constant $\L$.
Since the scalar curvature is constant, the Ricci tensor is proportional to the metric, which means that the solution is an Einstein manifold. In Euclidean signature the solution with positive $\L$ is the sphere $S^{d+1}$, the one with negative $\L$ is the hyperbolic space $H^{d+1}$, while the one with $\L=0$ is the flat space $R^{d+1}$. In Minkowskian signature, the maximally symmetric solution with positive $\L$ is called the de-Sitter (dS$_{d+1}$) spacetime, the one with negative $\L$ is the anti-de-Sitter ($AdS_{d+1}$) and the one with $\L=0$ is the Minkowski space.

The spaces above can be defined from a $d+2$ dimensional flat space via a quadratic constraint. We are interested in the $AdS$ case, which can be defined from $R^{2,d}$ space \cite{st}
\be ds^2=-dX_0^2-dX_{d+1}^2+\sum_{i=1}^d dX_i^2 \label{m2}\ee
as a hyperboloid of radius L via the condition:
\be X_0^2+X_{d+1}^2-\sum_{i=1}^d X_i^2=L^2 \label{c1}\ee

It is obvious from this condition that $AdS_{d+1}$ has isometry group SO(2,d) and it is homogeneous and isotropic. The constraint (\ref{c1}) can be solved using the following parametrization, \cite{st}:
\begin{align} X_0=L cosh(\rho) cos(\tau)\sp X_{p+1}=L cosh(\rho) sin(\tau)
\notag\\
X_i=L sinh(\rho) \Omega_i\sp \left(i=1,2,...,d \sp \Sigma_{i=1}^d \Omega_i^2=1\right) \label{p1}\end{align}
Substituting (\ref{p1}) into (\ref{m2}) we obtain the $AdS_{d+1}$ metric:
\be ds^2=L^2(-cosh^2(\rho)d\tau^2+d\rho^2+sinh^2(\rho)d\Omega_{p-1}^2) \label{m3}\ee
where $d\Omega_{p-1}^2$ is the line element of the sphere $S^{p-1}$. Taking $\rho\in \Re^+, \tau\in [0,2\pi)$ we cover the hyperboloid exactly once. Because of that these coordinates are called global. Since $\tau$ is periodic we have closed time-like curves. We can obtain a causal space-time by taking the universal cover, which means $\tau\in \Re$.

There is another useful set of coordinates defined as, \cite{Mald},\cite{st}:
\begin{align} X_0={u\over 2}\left[1+{1\over u^2}(L^2+\vec{x}^2-t^2)\right]\sp X_i={L x^i\over u}
\notag\\
X_{d}={u\over 2}\left[1-{1\over u^2}(L^2-\vec{x}^2+t^2)\right]\sp X_{d+1}={Lt\over u}\end{align}

which brings the metric to the form
\be ds^2={L^2\over u^2}(-dt^2+du^2+d \vec{x}^2) \ee
These coordinates are called Poincar\'e coordinates and they cover half the hyperboloid. The Poincar\'e symmetry of $t\sp x^i$ coordinates is now obvious, as well as the scale invariance: $t\to \l t\sp u\to \l u\sp x^i\to \l x^i$. In this coordinate system there is a boundary at $u=0$. This is the coordinate system we use in this thesis.

Finally there is another metric, also called Poincar\'e metric, related to the previous one by the transformation $r=L^2/u$:
\be ds^2=L^2\left[{dr^2\over r^2}+r^2(-dt^2+d\vec{x}^2)\right]\ee
The boundary is located at $r=\infty$.

\section{On Hyperscaling Violating metrics\label{HVM}}

Holography can be generalized to geometries which are not asymptotically AdS in the IR. Such a generalization is very useful in Condensed Matter (CM) physics. The category of Lifshitz theories is prime example. Such theories have a homogeneous scale invariant space characterized by the Lifshitz exponent, z:
\be t\to \lambda^z t\sp x_i\to \lambda x_i\ee
where t is the time component and $x^i$ are the spatial components of the space-time.

The gravitational dual of a (d+1)-dimensional Lifshitz field theory can be defined on a metric with the same symmetry:
\be ds_{d+2}^2=-{dt^2\over r^{2z}}+{dr^2+dR_d^2\over r^2}\ee
where $dR_d^2=\Sigma_{i=1}^d dx_i^2$ is the d-dimensional Euclidean metric.

The characteristic of this space-time is the anisotropy between time and space. This geometry is not a solution to a pure cosmological Einstein gravity, because there is nothing to produce such an anisotropy. One needs to couple gravity to other fields in order to obtain a Lifshitz metric. The Lifshitz geometry arises from an Einstein-Maxwell theory, where gravity is coupled to a gauge field.

There is a more general class of geometries which can be reached by also including a scalar field: they are solutions of an Einstein-Maxwell-Dilaton (EMD) theory. These metrics are conformal to Lifshitz and feature an extra parameter $\theta$, called the "Hyperscaling Violation" exponent:
\be ds_{d+2}^2=r^{2\theta\over d}(-{dt^2\over r^{2z}}+{dr^2+dR_d^2\over r^2}) \label{L}\ee
This is the most general metric with homogeneous space and is invariant under the transformation:

\be t\to \lambda^z t\sp r\to \lambda r\sp x_i\to \lambda x_i \sp ds_{d+2}\to \lambda^{\theta\over d} ds_{d+2} \label{HV}\ee

Obviously, when $\theta$ is non-zero the proper distance is not invariant under the scaling, which indicates violation of hyperscaling in the dual (d+1)-dimensional theory, \cite{HSS}. In a field theory dual to the geometry (\ref{L}) ($\theta=0$) the thermal entropy scales at low temperatures as $T^{d\over z}$ but in a field theory dual to (\ref{HV}) ($\theta\ne 0$) the entropy scales as $T^{d-\theta\over z}$. Therefore a theory with non-zero $\theta$ has effectively $d_{eff}=d-\theta$ dimensions at low temperature.

\subsection{Properties}
The hyperscaling violating metric is the most general metric with homogeneous space and is extremal (zero temperature),
\be ds_{d+2}^2=r^{2\theta\over d}(-{dt^2\over r^{2z}}+{dr^2+dR_d^2\over r^2})\ee

The IR is located either at $r=0$ or $r=\infty$, depending on the values of $\theta$, z, d. According to appendix B of \cite{gk} for the IR to be well defined we require:
\be (\theta-d)(\theta-dz)>0\label{IR}\ee
The IR is located at infinity if:
\be \theta<d\sp \theta<dz\ee
and is located at $r=0$ if:
\be \theta>d\sp \theta>dz\ee
In addition perturbations (appendix \ref{Perturbations}) with a mode $d+z-\theta$ in general have finite temperature according to appendix B of \cite{gk}. These modes should vanish in the UV, therefore in the IR we also require:
\bsea d+z-\theta>0 \sp r\xrightarrow[IR]{}\infty\\
 d+z-\theta< 0\sp r\xrightarrow[IR]{}0\label{IR2}\esea

The sign of $\theta$ determines the location of the singularity. The Ricci and Kretschmann scalars are:
\be R\propto r^{-2\theta\over d} \sp R_{\m\n\r\s}R^{\m\n\r\s}\propto r^{-4\theta\over d}\ee
Therefore if $\theta=0$ there are no singularities, if $\theta>0$ the singularity is located at $r=0$ and if $\theta<0$ it is located at infinity.

\section{Equations of motion and ansatz \label{appA}}
We start with the action:

\be S=M^{2}\int \ud^{4}x\le[ \sqrt{-g}\left(R-\frac12(\partial \phi)^2+V(\phi)-{1\over 4}Z(\phi) F^2\right) - W(\phi)F_{\mu\nu}\tilde F^{\m\n}  \ri]\sp \tilde{F}^{\m\n}={1\over 2}\e ^{\m\n\r\s}F_{\r\s} \ee

By varying the metric $g_{\m\n}$, the gauge field $A_\m$ and the scalar field $\phi$ we obtain respectively:

\bsea 	R_{\mu\nu}-{R+V(\phi)\over 2}g_{\mu\nu}&=& \half\partial_\mu\phi\partial_\nu\phi-\frac{g_{\mu\nu}}4\le(\partial\phi\ri)^2+{Z(\phi)\over
2}\le[F^{\;\rho}_\mu\ F_{\nu\rho}-\frac{g_{\mu\nu}}4 F^2\ri]\slabel{EEAA}  \\ 	 0&=&\nabla_\mu\le(Z(\phi)F^{\mu\nu}+{W(\phi)\over \sqrt{-g}}\tilde F^{\m\n}\ri)\,,
 \\ 	 \square{\phi}&=&\frac{Z'(\phi)}4F^2-V'(\phi)+\frac{W'(\phi)}{4\sqrt{-g}}\tilde{F}^2,\slabel{Dilaton_eq}  	 
 \label{14c}\esea
where \be \tilde{F}^2=F_{\mu\nu}\tilde F^{\m\n}
\ee

We will use the following radial ansatz for the metric and the scalar field
\be
 	\ud s^2 = -D(r)\ud t^2 + B(r) \ud r^2 +C(r)\le(dx_1^2+dx_2^2\ri)\sp \phi(r)
	\label{3} \ee
while for the gauge field components
 \be A_t(r)\sp A_r=0\sp A_1={h\over 2}x_2\sp A_2=-{h\over 2}x_1\sp h=constant \label{4}\ee

Substituting (\ref{3}),(\ref{4}) into (\ref{14c}) we obtain 4 Einstein equations with only 3 linearly independent (the last 2 equations are identical):

\be \begin{split}& {C'\over C}\left(2{B'\over B}+{C'\over C}\right) =-2BV+ h^2Z{B\over C^2}+{Z\over D}A_t'^2+\phi'^2+4{C''\over C}\\
& \phi'^2+2BV= h^2Z{B\over C^2}+{Z\over D}A_t'^2+{C'\over C}\left({C'\over C}+2{D'\over D}\right) \\
& 2{C''\over C}+2{D''\over D}+\phi'^2  = 2BV+h^2Z{B\over C^2}+{Z\over D}A_t'^2+{C'\over C}\left({B'\over B}+{C'\over C}-{D'\over D}\right)+{D'\over D}\left({B'\over B}+{D'\over D}\right)
\\
& 2{C''\over C}+2{D''\over D}+\phi'^2= 2BV+h^2Z{B\over C^2}+{Z\over D}A_t'^2+{C'\over C}\left({B'\over B}+{C'\over C}-{D'\over D}\right)+{D'\over D}\left({B'\over B}+{D'\over D}\right)\end{split}
\label{EE0}\ee
as well as the gauge field equation (which has been integrated):

\be A_t'={(q+hW)\over Z}{ \sqrt{DB}\over C}\sp q,h={\rm constant}
 \label{GFE}\ee
 and finally the scalar field equation:
\be \phi''+\phi'({C'\over C}+{D'\over 2D}-{B'\over 2B})=-B\partial_\phi V+{h\sqrt{BD}A_t'\over CD}\partial_\phi W+\left({h^2B\over 2C^2}-{A_t'^2\over 2D}\right)\partial_\phi Z \label{deq0}\ee
We can substitute $A_t'$ from (\ref{GFE}) into the Einstein and scalar equations of motion.

\bsea  \phi'^2+2{C''\over C} &=&\left({D'\over D}+{B'\over B}+{C'\over C}\right){C'\over C}\\
{B\over C^2}\left({(q+hW)^2 \over
Z^2}+h^2\right)Z&=&{D''\over D}-{C''\over C}+{1\over 2}\left({C'\over C}-{D'\over D}\right)\left({B'\over B}+{D'\over D}\right)\\
BV+{1\over 4}{B'\over B}{C'\over C}&=& {1\over 2}\left({D''\over D}+{C''\over C}\right)-{1\over 4}{D'\over D}
  \left({B'\over B}+{D'\over D}-3{C'\over C}\right)
 \label{EEA}\esea

\be \phi''+\phi'({C'\over C}+{D'\over 2D}-{B'\over 2B})={B\over 2C^2}\partial_\phi \left({(q+hW)^2\over Z}+h^2Z-2C^2V\right) \label{deq} \ee
In the running scalar case we are interested in hyperscaling violating solutions, therefore we use the following ansatz:
\be D(r)=r^{\theta-2z}\sp B(r)=B_0r^{\theta-2}\sp C(r)=r^{\theta-2}\sp
A_t=A_0 r^{\z-z}
\label{ansatz}\ee

\section{Null-Energy Condition\label{NEC}}

There are constraints on the exponents $z,\theta$ stemming from the null-energy condition. This condition states that the contraction of the stress-energy tensor, $T$ with a null-vector, $N$ (a vector with zero length, $N_\m N^\m=0$), must be non-negative. We can use $G_{\m\n}=T_{\m\n}$, where $G$ is the Einstein tensor to obtain the easier-to-handle expression:
\be T_{\m\n}N^\m N^\n\geq 0\Rightarrow G_{\m\n}N^\m N^\n\geq 0 \ee
Using (\ref{ansatz}) we obtain the conditions:
\be (2 - 2 z + \theta) (\theta-2)\geq 0 \sp (z-1) (2 + z - \theta)\geq 0\label{nec}\ee
The allowed values for $z,\theta$ are given in the left figure of \ref{NECPlot}. Adding the conditions for a well-defined IR (see appendix \ref{HVM}) we obtain the right figure of \ref{NECPlot}.

\begin{figure}[H]
\centering
\includegraphics[width=0.49\textwidth]{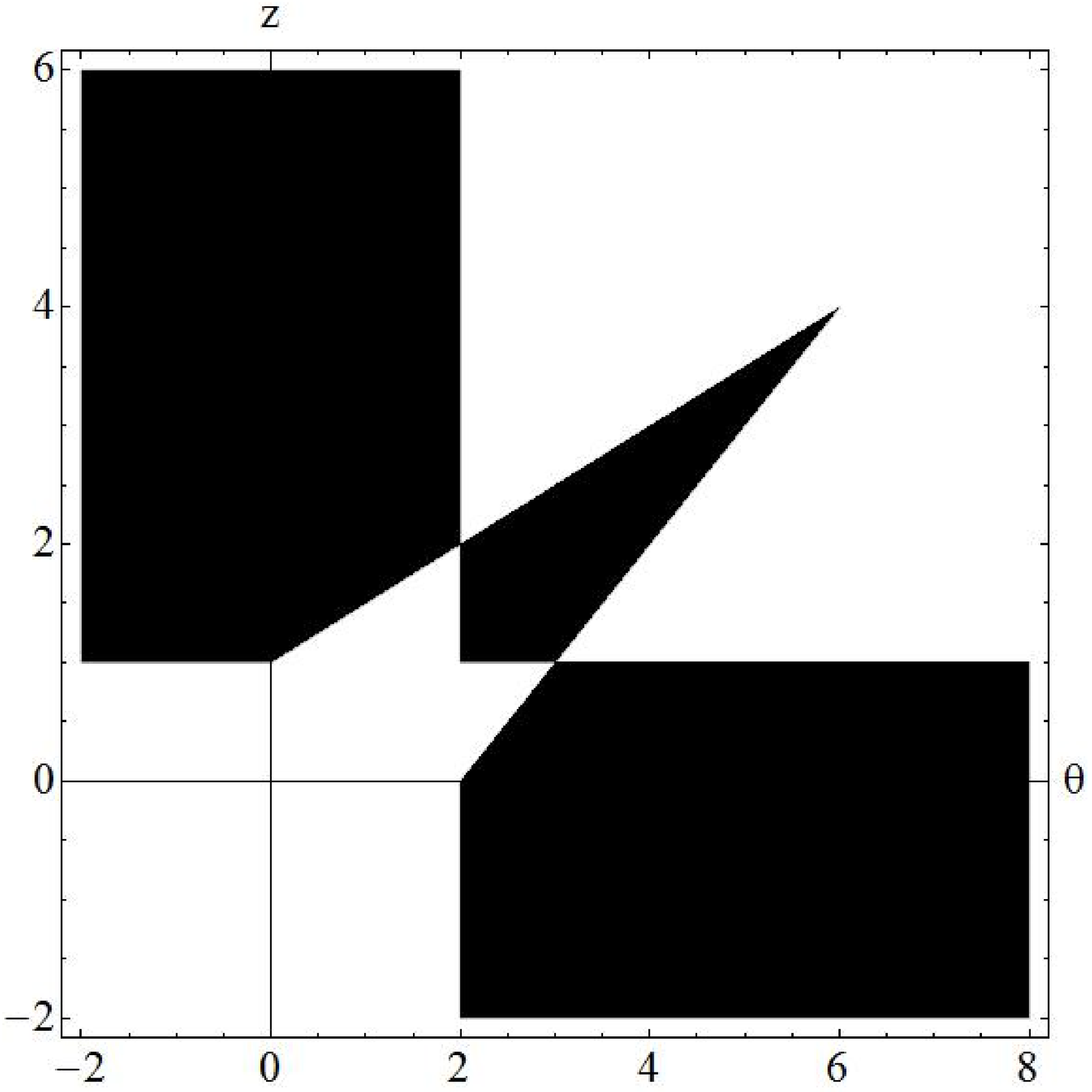}
\includegraphics[width=0.49\textwidth]{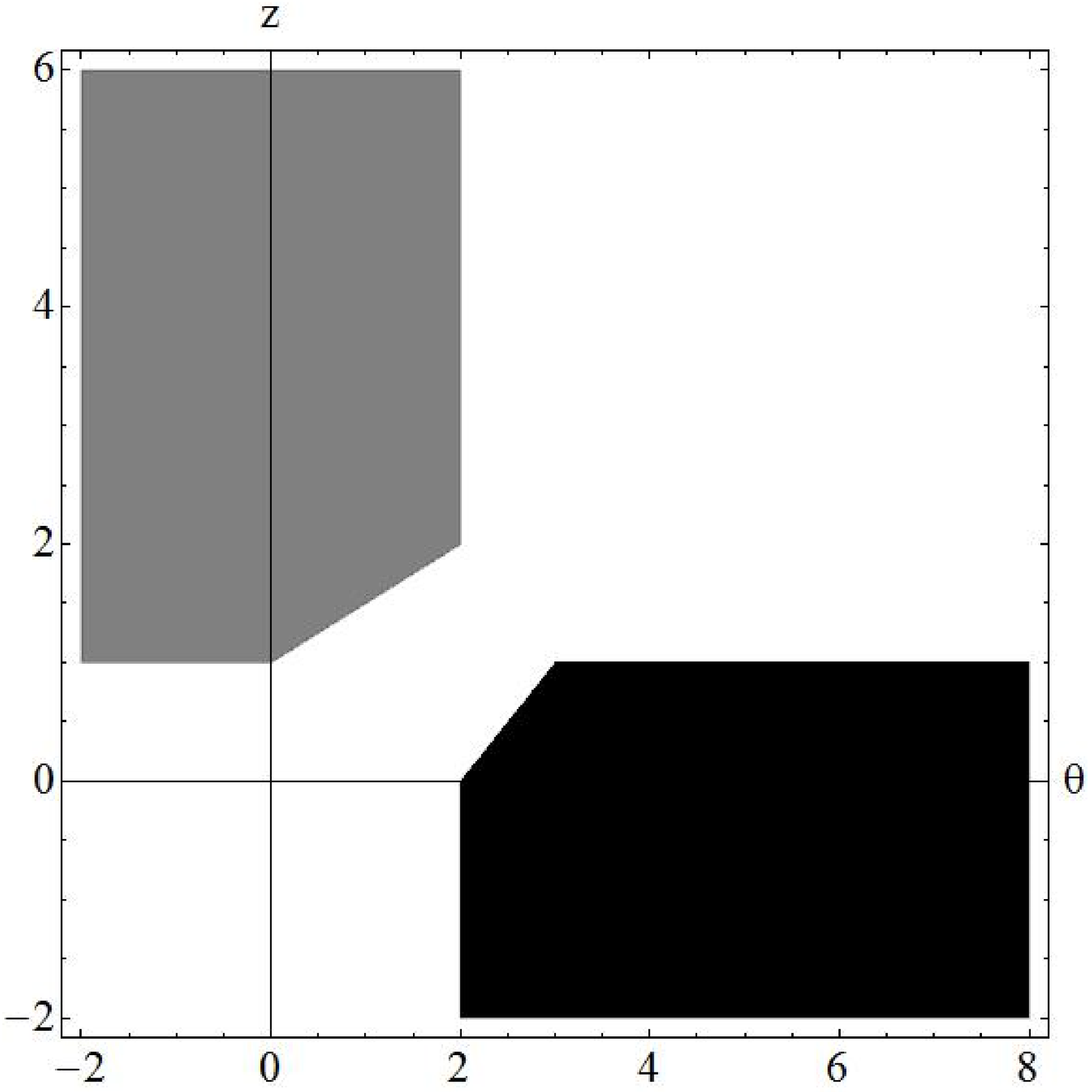}
\caption{\small Left figure: Imposing only the null-energy condition\textsuperscript{\protect\ref{nec}} the allowed values of $z,\theta$ are within the black area.
Right figure: Additionally we impose the conditions for well-defined IR. Within the gray area the IR is located at $r\xrightarrow[IR]{}\infty$ while in the black area the IR is located at $r\xrightarrow[IR]{}0$.}
\label{NECPlot}
\end{figure}

\section{Constant scalar solutions \label{appB}}

Setting $\phi=\phi_\star$ in (\ref{EEA}) we obtain:

\bsea  2{C''\over C} &=&\left({D'\over D}+{B'\over B}+{C'\over C}\right){C'\over C}\\
{B\over C^2}\left({(q+hW_\star)^2 \over
Z_\star^2}+h^2\right)Z_\star&=&{D''\over D}-{C''\over C}+{1\over 2}\left({C'\over C}-{D'\over D}\right)\left({B'\over B}+{D'\over D}\right)\\
BV_\star+{1\over 4}{B'\over B}{C'\over C}&=& {1\over 2}\left({D''\over D}+{C''\over C}\right)-{1\over 4}{D'\over D}
  \left({B'\over B}+{D'\over D}-3{C'\over C}\right)\label{CD0}\}\esea

While from the scalar equation (\ref{deq}) we have
\be \partial_\phi \left({(q+hW)^2\over Z}+h^2Z-2C^2V\right)\bigg|_{\phi=\phi_\star}=0 \label{CD}\ee

Where $V_\star,Z_\star,W_\star$ is the value of $V,Z,W$ at $\phi=\phi_\star$ respectively.

\subsection{Neutral solution}

We will look for solutions with $q=0=h$. This case is equivalent to a pure Einstein-Dilaton theory and has already been studied in the past\textsuperscript{\cite{gk}}. Setting $q=0=h$ and $\phi=\phi_\star$ in (\ref{CD}) we obtain:

\be \partial_\phi V(\phi_\star)=0\ee
Therefore the scalar settles to a constant $\phi_\star$ which extremizes the value of the scalar potential V.

Setting $q=0=h$ and $\phi=\phi_\star$ in (\ref{EEA}), the Einstein equations become:

\bsea  2{C''\over C} &=&\left({D'\over D}+{B'\over B}+{C'\over C}\right){C'\over C}\\
{C''\over C}-{D''\over D}&=&{1\over 2}\left({C'\over C}-{D'\over D}\right)\left({B'\over B}+{D'\over D}\right)\\
BV_\star+{1\over 4}{B'\over B}{C'\over C}&=& {1\over 2}\left({D''\over D}+{C''\over C}\right)-{1\over 4}{D'\over D}
  \left({B'\over B}+{D'\over D}-3{C'\over C}\right)
\label{CD1}\esea
Looking for solutions of the form:
\be ds^2=L^2 r^{\theta-2}(-dt^2+dr^2+dx_idx^i)\ee
we obtain
\be \theta (\theta-2)=0\sp L^2 V_\star r^\theta=(\theta-2)(\theta-3)\ee
from which we obtain
\be \theta=0, L^2={6\over V_\star}\sp V_\star\ne 0\ee
\be \theta=2\sp V_\star= 0\ee

\subsection{Dyonic solution}

We look for solutions with $qh\ne 0$ in general.
We rewrite (\ref{CD})
\bsea \left({D'\over D}+{B'\over B}+{C'\over C}\right){C'\over C}&=& 2{C''\over C} \\
{1\over 2}\left({C'\over C}-{D'\over D}\right)\left({B'\over B}+{D'\over D}\right)&=&{B\over C^2}E_\star+{C''\over C}-{D''\over D} \\
 {C'\over C}\left({C'\over
2C}+{D'\over D}\right)+{B\over 2C^2}E_\star&=& BV_\star \label{EE2}\esea
where \be E_\star=h^2Z_\star+{(q+hW_\star)^2\over Z_\star}\ee
From the equation (\ref{CD}) we obtain
\be
C^2={E'_\star\over 2 V'_\star}={\rm constant}
\label{24}\ee
Setting $C'=C''=0$ in (\ref{EE2}) we are left with
\bsea
{D''\over D}-{1\over 2}{D'\over D}\left({B'\over B}+{D'\over D}\right)&=&{B\over C^2}E_\star \slabel{25}\\
 2C^2V_\star &=& E_\star \slabel{26}\esea
From (\ref{24}) and (\ref{26}) we obtain
\be V'_{eff}(\phi_\star)=V'_\star-{E'_\star\over E_\star}V_\star\label{27}\ee
The only equation that remains is (\ref{25}) which using (\ref{26}) can be written as
\be
{D''\over D}-{1\over 2}{D'\over D}\left({B'\over B}+{D'\over D}\right)=2V_\star B
\ee
We use the gauge freedom stemming from the radial ansatz to set:
\be
D={L ^2\over r^2}f(r)\sp B={L ^2\over r^2f(r)}\sp L^2={1\over V_*}
\ee
The equation becomes
$$
r^2f''-2rf'+2f-2=0
$$
with the general solution
\be
f(r)=1+K_1 r+K_2r^2
\ee

The metric is therefore:

\be ds^2=L^2\left(-{f(r)\over r^2}dt^2+{dr^2\over f(r)r^2}\right)+C_0 dx_idx^i\ee
With the transformation
\be\rho={2\arctan{f'(r)\over \sqrt{-\Delta}}\over \sqrt{-\Delta}}\ee the metric becomes:

\be ds^2=L^2G(\rho)(-dt^2+d\rho^2)\sp  G(\rho)=       {      -\Delta
 \over 16 \cos^2({1\over 2}\sqrt{-\Delta}\rho+\arctan({\sqrt{-\Delta} \over K_1}))}\ee
 We transform again the radial coordinate and rescale the time coordinate as:

 \be u={1\over 2}\sqrt{-\Delta}\rho+\arctan({\sqrt{-\Delta} \over K_1})\sp t={2\tau\over \sqrt{-\Delta}}\sp \tilde{L}={L/\sqrt{2}}\ee
 to obtain:
\be ds^2={\tilde{L}^2\over \cos^2 u}(-d\tau^2+du^2)\ee
which is AdS$_2$. Since the transformation involved $\sqrt{-\Delta}$ it becomes singular when $\Delta>0$.

Plugging the solution into (\ref{GFE}) we obtain:

\be A_t=\m-{Q\over r}\sp Q^{-2}={V_\star Z_\star\over 2}\left(1+\left({hZ_\star\over q+hW_\star}\right)^2\right)\sp \m={\rm constant}\ee

\section{Running scalar in the IR \label{appC}}

We assume that the scalar field runs logarithmically in the IR and the coupling functions scale as:

\be V(\phi)=V_0 e^{-\d\phi}\sp Z(\phi)=Z_0e^{\g\phi}\sp W(\phi)=W_0 e^{\chi\phi}\sp e^{\phi}=e^{\phi_0}r^a \label{s}\ee

Using the ansatz (\ref{ansatz}) with the Einstein equations (\ref{EEA}) we obtain:

\bsea k_1 &=& a^2 \\
 k_2 &=& B_0 r^{4-\theta}\left(h^2Z+{(q+hW)^2\over Z}\right)\\
 k_3 &=& B_0 V r^\theta\esea

 where
 \be k_1=(\theta-2)(\theta-2z+2)\sp k_2=2(1-z)(\theta-z-2)\sp k_3=(\theta-z-2)(\theta-z-1)\ee

 also from the gauge field equation (\ref{GFE}) we obtain:
 \be A_0(\zeta-z)r^{\zeta-2} = \sqrt{B_0}{(q+hW)\over Z} \ee

We ignore the scalar equation because it always satisfied if the above equations are satisfied.

\subsection{Solutions without intrinsic P-violation}

In this section we ignore the parity violating term $F\wedge F$ of the action. We rewrite the above equations setting $W=0$.

The Einstein equations are the following:
\bsea k_1 &=& a^2 \\
 k_2 &=& B_0 r^{4-\theta}\left(h^2Z+{q^2\over Z}\right)\slabel{Z}\\
 k_3 &=& B_0 V r^\theta\slabel{V}\esea
and the gauge field equation is:
 \be A_0(\zeta-z)r^{\zeta-2} = \sqrt{B_0}{q\over Z} \ee

In equation (\ref{Z}) there are two terms on the right hand side which depend on different powers of r (because of the implicit r-dependence of Z). We have to consider cases depending on which one of those terms is leading in the IR limit. For that reason we distinguish 3 cases regarding Z. From (\ref{s}) we see that Z scales as $Z\sim r^{a\g}$. Depending on the sign of $a\g$ and whether the IR is located at $0$ or $\infty$ we have the following cases:

\begin{enumerate}[I]

\item $Z\to 0$: Then $a\g>0$ ($a\g<0$) for $r\to 0$ ($r\to\infty$). In this case in leading order $h^2Z+{q^2\over Z}\to {q^2\over Z}$.

 This is the electric solution. This case can also be reached by setting $h=0$ and the constraint on the sign of $a\g$ is evaded.

\item $Z\to\infty$: Then $a\g<0$ ($a\g>0$) for $r\to 0$ ($r\to\infty$). In this case in leading order $h^2Z+{q^2\over Z}\to h^2Z$

 This is the magnetic solution. This case can also be reached by setting $q=0$ and the constraint on the sign of $a\g$ is evaded.

\item $Z\to Z_0e^{\g\phi_0}$: Then $a\g=0$. In this case both terms are of the same order $h^2Z+{q^2\over Z}\to h^2Z_0e^{\g\phi_0}+{q^2\over Z_0}e^{-\g\phi_0}$

This case contains both electric and magnetic fields, however because the two terms are of the same order only if they are both constant we have the constraint $a\g=0$.
\end{enumerate}

Next we notice that in (\ref{Z}), (\ref{V}) the left hand side is a constant and the right hand side is a function of r. The right hand sides can be leading or subleading in the IR. We have to further distinguish 4 subcases regarding the constants $k_2,k_3$

\begin{enumerate}[a]
\item $k_2k_3\ne 0$:

In this subcase all the terms are leading.

\item $k_2\ne 0\sp k_3=0$:

In this subcase the right hand side of (\ref{V}) is subleading. This corresponds to solutions with subleading scalar potential ($V\to 0$).

\item $k_2= 0\sp k_3\ne 0$:

In this subcase the right hand side of (\ref{Z}) is subleading. This corresponds to neutral solutions.

\item $k_2=0=k_3$:

In this subcase both right hand sides of (\ref{V}) and (\ref{Z}) are subleading. This corresponds to neutral solutions with subleading scalar potential ($V=0$). The geometry in this case does not depend on the parameters $\d,\g$ of the theory.

\end{enumerate}

\subsubsection{Electric solutions}

\begin{enumerate}[a]
\item $k_2k_3\ne 0$:

\be a={4\over \gamma+\delta}\sp \theta={4\d\over \g+\d}\sp z={3\d^2-\g^2-2\g\d-4\over \d^2-\g^2}\sp \zeta=\theta-2\ee
\be B_0={k_3\over V_0}e^{\d\phi_0}\sp q^2=2{2+\d(\g-\d)\over 2+\g(\g-\d)}Z_0V_0e^{(\g-\d)\phi_0}\sp A_0={q\sqrt{B_0}\over Z_0(\theta-z-2)}e^{-\g\phi_0}\ee

$$ a\g<0\sp r\xrightarrow[IR]{} \infty$$
$$a\g>0\sp r\xrightarrow[IR]{} 0$$

\item $k_2\ne 0\sp k_3=0$:

\be a={2\g\over 1+\ga^2}\sp \theta=2+{2\over 1+\g^2}\sp z=\theta-1\sp \zeta=\theta-2\ee

\be B_0q^2={4Z_0e^{\g\phi_0}\over 1+\gamma^2}\sp A_0=-{2e^{-{1\over 2}\g\phi_0}\over \sqrt{Z_0 (1+\g^2)}} \ee

$$ a\g<0\sp\theta<a\d\sp r\xrightarrow[IR]{} \infty$$
$$ a\g>0\sp\theta>a\d\sp r\xrightarrow[IR]{} 0$$
\item $k_2= 0\sp k_3\ne 0$:

\be a={2\d\over \d^2-1}\sp \theta={2\d^2\over \d^2-1}\sp z=1\ee

\be B_0={2e^{\d\phi_0}\over V_0}{3-\delta^2\over (\d^2-1)^2}\ee

$$ a\g<0\sp\theta+a\g>4\sp r\xrightarrow[IR]{} \infty$$
$$ a\g>0\sp\theta+a\g<4\sp r\xrightarrow[IR]{} \infty$$

\item $k_2=0=k_3$:

\be a^2=3\sp z=1\sp \theta=3\ee

$$a\g<0\sp\theta<a\d\sp\theta+a\g>4\sp r\xrightarrow[IR]{} \infty$$
$$a\g>0\sp\theta>a\d\sp\theta+a\g<4\sp r\xrightarrow[IR]{} 0$$

\end{enumerate}

\subsubsection{Magnetic solutions}

\begin{enumerate}[a]
\item $k_2k_3\ne 0$:

The same as Ia with $q^2\to h^2 Z_0^2\sp \gamma\to -\gamma$

$$a\g>0\sp r\xrightarrow[IR]{} \infty$$
$$a\g<0\sp r\xrightarrow[IR]{} 0$$

\item $k_2\ne 0\sp k_3=0$:

The same as Ib with $q^2\to h^2 Z_0^2\sp \gamma\to -\gamma$

$$a\g>0\sp \theta<a\d\sp r\xrightarrow[IR]{} \infty$$
$$a\g<0\sp \theta>a\d\sp r\xrightarrow[IR]{} 0$$

\item $k_2= 0\sp k_3\ne 0$:

The same as Ic with $\gamma\to -\gamma$

$$a\g>0\sp \theta>4+a\gamma\sp r\xrightarrow[IR]{} \infty$$
$$a\g<0\sp \theta<4+a\gamma\sp r\xrightarrow[IR]{} 0$$
\item $k_2=0=k_3$:

The same as Id with $\gamma\to -\gamma$
\item $a^2=z(4-z)\sp \theta=2+z\sp a\g=0$

$$a\g<0\sp\theta<a\d\sp\theta>4+a\g\sp r\xrightarrow[IR]{} \infty$$
$$a\g>0\sp\theta>a\d\sp\theta<4+a\g\sp r\xrightarrow[IR]{} 0$$

\end{enumerate}

\subsubsection{Dyonic solutions}

\begin{enumerate}[a]
\item $k_2k_3\ne 0$:

Same geometry as Ia with $q^2\to q^2+h^2 Z_0^2\sp \gamma=0$

Same gauge field as Ia with $\g=0$

$$ \g=0$$

\item $k_2\ne 0\sp k_3=0$:

Same geometry as Ib with $q^2\to q^2+h^2 Z_0^2\sp \gamma=0$

Same gauge field as Ia with $\g=0$

\be \g=0\sp\theta<a\d\ee

\item $k_2= 0\sp k_3\ne 0$:

Same geometry as Ic with $a=0$

\item $k_2=0=k_3$:

Same as Id

\be a\g=0\sp\theta<a\d\sp\theta>4\ee
\end{enumerate}

\subsection{Solutions with intrinsic P-violation}

We distinguish 3 cases for W in the IR limit:

\begin{itemize}
\item $\lim W=0$ Then $a\chi>0$ ($a\chi<0$) for $r\to 0$ ($r\to\infty$). In the leading order solution ${q+hW\over Z}\to {q\over Z}$

We obtain the same leading order solutions as the previous section.

\item $\lim W=W_0$ Then $a\chi=0$. In the leading order solution ${q+hW\over Z}\to {q+hW_0\over Z}$

This case contains both an electric and a magnetic field. We obtain the same leading order solutions as the previous case with the difference $q\to q+hW_0$.

\item $\lim W=\infty$ Then $a\chi<0$ ($a\chi>0$) for $r\to 0$ ($r\to\infty$). In the leading order solution ${q+hW\over Z}\to {hW\over Z}$

We can also arrive at this case by setting $q=0$ to avoid the constraints on the sign of $a\chi$.
\end{itemize}

For $W\to \infty$ we rewrite the equations:

\bsea k_1 &=& a^2 \\
 k_2 &=& B_0h^2 r^{4-\theta}\left(Z+{W^2\over Z}\right)\\
 k_3 &=& B_0 V r^\theta\esea

and the gauge field equation:
 \be A_0(\zeta-z)r^{\zeta-2} = \sqrt{B_0}h{W\over Z} \ee

 We distinguish 3 cases regarding $W/Z$:

 \begin{enumerate}[I]
 \item ${W\over Z}\to 0$ Then $a(\chi-\g)>0$ ($a(\chi-\g)<0$) for $r\to 0$ ($r\to\infty$). In the leading order solution $Z+{W^2\over Z}\to Z$.

 In this case we obtain again the magnetic solutions of the previous section.

  \item ${W\over Z}\to {W_0\over Z_0}$ Then $a(\chi-\g)=0$. In the leading order solution $Z+{W^2\over Z}\to Z(1+{W_0^2\over Z_0^2})$

  	The same the previous case, but with $Z_0\to Z_0+{W_0^2\over Z_0}\sp a\chi=a\g$

 \item ${W\over Z}\to \infty$ Then $a(\chi-\g)<0$ ($a(\chi-\g)>0$) for $r\to 0$ ($r\to\infty$). In the leading order solution $Z+{W^2\over Z}\to {W^2\over Z}$.

 In this case we have new solutions.

 \end{enumerate}

 We further distinguish 4 subcases regarding the constants $k_2,k_3$

 \begin{enumerate}[a]
 \item $k_2k_3\ne 0$:

 This subcase gives electric/magnetic solutions.

 \item $k_2\ne 0\sp k_3=0$:

 This subcase corresponds to solutions with subleading scalar potential ($V=0$).

 \item $k_2= 0\sp k_3\ne 0$:

 This subcase corresponds to neutral solutions.

 \item $k_2=0=k_3$:

 This subcase corresponds to neutral solutions with subleading scalar potential ($V=0$).

 \end{enumerate}

  \begin{enumerate}[a]
   \item $k_2k_3\ne 0$:
   \be a={4\over \g+\d-2\chi}\sp \theta={4\d\over \g+\d-2\chi}\sp z={3\d^2-(\g-2\chi)^2-2\d(\g-2\chi)-4\over \d^2-(\g-2\chi)^2}\sp \zeta=2{\d-\g\over \g+\d-2\chi}\ee

   \be B_0={k_3\over V_0}e^{\d\phi_0}\sp h^2={k_2\over k_3}{Z_0V_0\over W_0^2}e^{(\g-\d-2\chi)\phi_0}\sp A_0=\sqrt{B_0}{hW_0\over (\zeta-z)Z_0}e^{(\chi-\gamma)\phi_0}\ee

$$ a\chi>a\g\sp r\xrightarrow[IR]{} \infty$$
   $$ a\chi<a\g\sp r\xrightarrow[IR]{} 0$$

   \item $k_2\ne 0\sp k_3=0$:
   \be a={2(\g-2\chi)\over 1+(\g-2\chi)^2}\sp \theta=2+{2\over 1+(\g-2\chi)^2}\sp z=\theta-1\sp \zeta=2(\chi-\gamma){1+2(\g-2\chi)^2\over 1+(\g-2\chi)^2}\ee

   \be B_0h^2={2(\theta-2)Z_0\over W_0^2} e^{(\g-2\chi)\phi_0}\sp A_0=\sqrt{B_0}{hW_0\over (\zeta-z)Z_0}e^{(\chi-\gamma)\phi_0}\ee

$$ a\chi>a\g\sp \theta<a\d\sp r\xrightarrow[IR]{} \infty$$
   $$ a\chi<a\g\sp \theta>a\d\sp r\xrightarrow[IR]{} 0$$

   \item $k_2= 0\sp k_3\ne 0$:

   The geometry is the same as Ic of the parity preserving case.

      $$ a\chi>a\g\sp \theta>4+a(2\chi-\g)\sp r\xrightarrow[IR]{} \infty$$
         $$ a\chi<a\g\sp \theta<4+a(2\chi-\g)\sp r\xrightarrow[IR]{} 0$$

   \item $k_2=0=k_3$:

   The geometry is the same as Id of the parity preserving case.

      $$  a\chi>a\g\sp \theta<a\d\sp \theta>4+a(2\chi-\g)\sp r\xrightarrow[IR]{} \infty$$
         $$  a\chi<a\g\sp \theta>a\d\sp \theta<4+a(2\chi-\g)\sp r\xrightarrow[IR]{} 0$$
   \end{enumerate}

\section{Solutions with subleading scalar potential ($V\to 0$)}\label{V0}

In this section we present the solutions in which the scalar potential $V$ is subleading in the IR. The solutions are problematic because the IR is usually not well-defined and important parameters, such as $B_0$, cannot be fixed.

\begin{enumerate}[I]
\item{Neutral solution \label{Id}}

This case corresponds to an Einstein-Dilaton theory with $V=0$.
The solution is conformally flat:

\be ds^2=r\left(-dt^2+B_0dr^2+dx_idx^i\right)\sp e^\phi=e^{\phi_0} r^a\ee

with $B_0$ undetermined and $a^2=3$. The IR located at $r=0$.

\item{Magnetic solution at zero charge density\label{IIb}}

The IR is not well-defined and the solution is valid only for $r\to 0$.

\be ds^2=-r^{2-\theta}dt^2+r^{\theta-2}\left(B_0 dr^2+dx_idx^i\right)\sp e^\phi=e^{\phi_0}r^a\ee

\be a=-{2\g\over 1+\ga^2}\sp \theta=2+{2\over 1+\g^2}\sp z=\theta-1\sp \zeta=\theta-2\ee

\be B_0Z_0h^2={4e^{-\g\phi_0}\over 1+\gamma^2}\ee

Since $\theta$ is always positive, the singularity is located at $r=0$. For $\g\to\infty$ we arrive at a flat, neutral geometry.

\item{Electric solution at finite charge density\label{Ib}}

The IR is not well-defined and the solution is valid only for $r\to 0$. This is the electric dual of \ref{IIb}. There is also a charge density:

\be A_t=\m+{A_0\over r}\sp A_0=-{2e^{-{1\over 2}\g\phi_0}\over \sqrt{Z_0 (1+\g^2)}} \ee

\item{Dyonic solution at finite charge density}

In the leading order solution there is both a finite charge density and a non-trivial magnetic field only if $Z$ is constant. In this case the scalar field must be constant $a=0$, and thus $\g$ is free.

\be ds^2=-{dt^2\over r^2}+r^2\left(B_0dr^2+dx_idx^i\right)\sp A_t=\m+{A_0\over r}\sp A_0=-{q\sqrt{B_0}\over Z_0 e^{\g\phi_0}}\ee

\be a=0\sp \theta=4\sp z=3\sp \zeta=2\ee

\be B_0\left(e^{\g\phi_0}h^2Z_0+{q^2\over Z_0}e^{-\g\phi_0}\right)=4\ee

The IR is not well-defined and the solution is valid only for $r\to 0$.

\item{Magnetic solution at finite charge density}

In this case $W$ is participating in the leading order solution and $V$ is subleading. The solution is valid only for $r\to 0$ and the IR is not well-defined. For $\chi\to\infty$ we obtain a flat neutral geometry with constant scalar, while for $\chi=\g/2$ the geometry is conformally flat.

\be ds^2=-r^{2-\theta}dt^2+r^{\theta-2}\left(B_0dr^2+dx_idx^i\right)\sp  e^{\phi}=e^{\phi_0}r^a\sp A_t=\m+A_0 r^{\zeta-z}\ee

\be a={2(\g-2\chi)\over 1+(\g-2\chi)^2}\sp \theta=2+{2\over 1+(\g-2\chi)^2}\sp z=\theta-1\sp \zeta=2(\chi-\gamma){1+2(\g-2\chi)^2\over 1+(\g-2\chi)^2}\ee

   \be B_0h^2={2(\theta-2)Z_0\over W_0^2} e^{(\g-2\chi)\phi_0}\sp A_0={1\over \zeta-z}\sqrt{2(\theta-2)\over Z_0 e^{\g\phi_0}}\ee

\end{enumerate}

\section{Linear Perturbations \label{Perturbations}}
We consider two kinds of perturbations. The first is an expansion in amplitudes of the perturbations. The second is an expansion in powers of r, where we calculate the first order corrections due to the subleading terms in the leading order solutions of appendix \ref{appC}.  We are mostly interested in perturbations that maintain zero temperature. In the second case we have an inhomogeneous linear 5x5 system of equations with the amplitudes as the unknown variables. The inhomogeinity comes from the subleading term and the amplitudes can be fixed in terms of the parameters of the leading order solution, after choosing a gauge. In the first case the system is homogeneous and has a non-trivial solution only if its determinant vanishes.

We perturb around the running scalar solutions using the following ansatz:

\be B(r)=B_0 r^{\theta-2}(1+B_1 r^{b_1})\sp C(r)=r^{\theta-2}(1+C_1 r^{c_1})\sp D(r)=r^{\theta-2z}(1+D_1 r^{d_1})\ee

\be \phi=\phi_0+a\log(r)+\Phi_1 r^{\phi_1}\sp A_t=\mu+r^f(A_0+A_1 r^{f_1})\ee

We plug this perturbation ansatz into the equations of motion and set $c_1=d_1=b_1=a_1=b$. There is a residual gauge freedom stemming from the radial ansatz. Choosing a gauge practically means fixing one of the perturbation amplitudes. We will not choose a gauge, as the calculations are very simple even without a fixed gauge. We have a 5x5 homogeneous system with the perturbation amplitudes as unknown variables. There are non-trivial solutions only when the determinant vanishes. The determinant can be factored into:
\be b(b-2-z+\theta)(b-b_+)(b-b_-)(f_1+f)\ee
where
\be b_++b_-=2+z-\theta\sp b_+b_-=-4 { (z-1) (1 + z - \theta) (2 + z - \theta)\over 2 z-2  - \theta}\ee

There are two pairs of conjugate modes each adding up to $2+z-\theta$, which should be the case in this coordinate system. There is always a marginal mode $b=0$ which only gives a shift of the constants of the leading order solution. These shifts can always be absorbed by rescaling $t\sp x^i$ and choosing a gauge. The mode $b=2+z-\theta$ corresponds to a finite temperature perturbation, however with some gauge choices it can maintain zero temperature according to appendix B of \cite{gk}. In the gauge $C_1=0$ this mode always gives the emblackening factor and has finite temperature:
$f(r)=1+D_1 r^{2+z-\theta}$
\be ds^2=-r^{\theta-2z}f(r)dt^2+r^{\theta}{B_0dr^2\over r^2f(r)}+r^{\theta-2}dx_idx^i\ee

For the non-universal modes $b_\pm$ we can see that for the neutral solutions ($z=1$) they are identical to the universal modes: $b_+=2+z-\theta\sp b_-=0$. However in all other cases they maintain zero temperature and one of them is relevant while the other is irrelevant, since always $b_+b_-<0$.

\subsection{Neutral hyperscaling violating solution}

We perturb around the solution \ref{Ic}.

\begin{itemize}

\item{$b=0$}: Marginal mode

\be B_1=\d\Phi_1\ee

\item{$b=2+z-\theta$}: Relevant mode with finite temperature

\be B_1 = - D_1-2C_1+2\d\Phi_1\ee

\end{itemize}

\subsection{Electric/Magnetic hyperscaling violating solutions}\label{P1}

We perturb around the solution \ref{IIa2}, the perturbations for \ref{IIa}, \ref{Ia} and \ref{IIIa} can be obtained by setting $\chi=\g$, $\chi=0$ and $\chi=0=\g$ respectively.
  \begin{itemize}
\item{$b=0$}: Marginal mode

 \be B_1 = \d \Phi_1 \sp C_1 = \Phi_1 {2 \chi - \g + \d\over 2}\ee

\item{$b=2+z-\theta$}: Relevant mode with finite temperature
 \be B_1=-D_1 + (\d + \g-2\chi) \Phi_1\sp C_1={1\over 2}(\d - \g+2\chi) \Phi_1\ee

\item{$b=b_\pm$}: One relevant and the other irrelevant.

The expressions for the amplitudes are too messy and we will not present them. This pair of modes shows a dynamical instability ($b_\pm$ become complex) when
$$(2 + z - \theta) ( 2 z-2 - \theta) (-20 + 2 z + 18 z^2 + 16 \theta - 19 z \theta + \theta^2)<0$$ However such values are forbidden by combining the null energy condition (\ref{nec}) and the conditions for well-defined IR (\ref{IR}), (\ref{IR2}). Therefore the solutions are not dynamically unstable.
   \end{itemize}

\subsection{Neutral AdS$_4$}

We perturb around the neutral constant scalar solution (\ref{CS1}) using the ansatz (the perturbations around this solution have also been studied in \cite{gk})
\be B(r)={L^2\over r^2}(1+B_1 r^{b_1})\sp C(r)={L^2\over r^2}(1+C_1 r^{c_1})\sp D(r)={L^2\over r^2}(1+D_1 r^{d_1})\ee

\be \phi=\phi_\star+\Phi_1 r^{a_1}\sp A_t=\mu+A_1 r^{f_1}\ee

The determinant factors into
\be c_1d_1f_1(f_1-1)(a_1^2-3a_1+L^2 V''_\star)[c_1(6-3d_1+b_1)+3d_1+b_1(2d_1-9)]\ee

Starting with the gauge field perturbations, there are 2 modes $f_1=0, f_1=1$ corresponding to chemical potential and charge density. The latter mode creates a constant finite flux.

Next, coming from the scalar field, there are 2 modes given by:
\be a_1-3a_1+L^2 V''_\star=0\to a_1^\pm={1\over 2}\left[3\pm\sqrt{9-4L^2 V_\star''}\right] \ee
The mode $a_1^+$ is always relevant, while $a_1^-$ is relevant when $V_\star''>0$. The solution is dynamically unstable when $4L^2V_\star''>9$.

For the modes coming from the metric:
Choosing the gauge $C_1=0$ we obtain $d_1=b_1=3$ with amplitudes $B_1=-D_1\sp C_1=\Phi_1=A_1=0$. This is the finite temperature perturbation.

\subsection{Electromagnetic AdS$_2\times R^2$}

We perturb around the constant scalar $AdS_2\times R^2$ solutions  (\ref{cs2}), (\ref{cs3}), (\ref{cs4}) using the ansatz

\be B(r)={L^2\over r^2}(1+B_1 r^{b_1})\sp C(r)=C_0(1+C_1 r^{c_1})\sp D(r)={L^2\over r^2 }(1+D_1 r^{d_1})\ee

\be \phi=\phi_\star+\Phi_1 r^{\phi_1}\sp A_t=\mu+r^{-1}(A_0+A_1 r^{f_1})\ee
In all 3 cases the perturbation modes are similar. By setting $h=0$ or $q=0$ one obtains the perturbation for the pure electric and pure magnetic case respectively.

The determinant factors into:

\be c_1 d_1(1 + c_1)(d_1-2) (f_1-1)P(a_1)\ee
where

\be P(a_1)= a_1^2-a_1
      +
    L^2 V_\star'' -{E_\star''\over E_\star}\to a_1^\pm={1\over 2}\pm \sqrt{{1\over 4}-L^2V_\star''+{E_\star''\over E_\star}}\ee
where $E_\star=(q+hW_\star)^2/Z_\star+h^2Z_\star$.

There are six modes $b=(-1,0,1,2,a_+,a_-)$ which pairwise sum to $1$ as required. The amplitudes for each case are given below.
\begin{itemize}
\item{$b=-1$}: Irrelevant mode
\bsea B_1 &=& {1\over 3}(4 C_1 + 3 D_1 - 2 L^2 \Phi_1 V_\star')\sp C_1= \Phi_1({E_\star''\over 2E_\star'}-{V_\star''-2V_\star\over 2 V_\star'})\notag\\
 A_1 &=&{q+hW_\star\over 3 LZV_\star'\sqrt{2E_\star}}\left[\left({V_\star'^2\over V_\star^2}-1+3{V_\star'\over V_\star}({Z_\star'\over Z_\star}-{hW_\star'\over q+hW_\star})\right)\Phi_1+3{V_\star'\over V_\star}D_1\right]\notag   \esea
\item{$b=0$}: Marginal mode
\be B_1 = -\Phi_1 L^2  V_\star'\sp C_1=0\sp A_1={L^2(q+hW_\star)\over 2 CZ_\star}\left(-D_1+\Phi_1\partial_\phi \log({VZ^2\over (q+hW)^2})\bigg|_{\phi=\phi_\star}\right)\ee

\item{$b=1$}: Relevant mode, finite temperature, shift of chemical potential
\be B_1=-D_1 - 2\Phi_1 L^2  V_\star'\ee
We have additionally $\Phi_1=0$ in all cases except \ref{cs2}.

\item{$b=2$}: Relevant mode
\be C_1=\Phi_1=0\sp A_1=(B_1+D_1){L(q+hW_\star)\over Z_\star\sqrt{2E_\star}}\ee

\item{$b=a_1^\pm$}:
\be B_1=-a_1^\pm D_1\sp C_1=\Phi_1=0\sp A_1=-D_1{(q+hW_\star) L\over Z_\star \sqrt{2E_\star}}\ee
These two modes come from the scalar field. The mode $a_1^+$ is always relevant, while $a_1^-$ is irrelevant if $L^2V_\star''<{E_\star''\over E_\star}$. There is a dynamical instability if $L^2V_\star''>{1\over 4}+{E_\star''\over E_\star}$.

\end{itemize}

\subsection{Corrections due to the PQ term}\label{PQp}

When $W$ is subleading in the IR, the leading order solutions are identical to the solutions presented in section \ref{W0}. In this section we present these solution along with the first order corrections coming from the PQ term.

\subsubsection{Neutral solution\label{Ic2}}

The leading order solution in this case is described by \ref{Ic}.
\be\begin{split} & ds^2=-D(r)dt^2+B(r)dr^2+C(r)dx_idx^i\sp e^\phi=e^{\phi_0+\Phi_1 r^b}r^{a}\sp A_t=\m+A_1 r^\beta \\
& D(r)=r^{\theta-2z}(1+D_1r^b)\sp B(r)=B_0r^{\theta-2}(1+B_1r^b)\sp C(r)=r^{\theta-2}(1+C_1r^b)\\
& a={2\d\over \d^2-1}\sp \theta={2\d^2\over \d^2-1}\sp z=1\sp b=4+(2\chi-\g-\d)a\sp \beta=1+(\chi-\g)a \\
& B_0={2e^{\d\phi_0}\over V_0}{3-\delta^2\over (\d^2-1)^2}\end{split}\ee

All the amplitudes $D_1,B_1,C_1,\Phi_1$ are fixed (after choosing a gauge) and are proportional to $h^2{W_0^2\over Z_0}$, while $A_1$ is proportional to $h{W_0\over Z_0}$. The region of validity of this solution is smaller than that of \ref{Ic} as we also require that $b,\beta$ be subleading in the IR.

\subsubsection{Magnetic solution at zero charge density}\label{IIa1}
The leading order solution in this case is the same as \ref{IIa}.
\be\begin{split} & ds^2=-D(r)dt^2+B(r)dr^2+C(r)dx_idx^i\sp  e^{\phi}=e^{\phi_0+\Phi_1 r^b}r^{a}\sp A_t=A_t=\m+A_1 r^\beta\\
& D(r)=r^{\theta-2z}(1+D_1r^b)\sp B(r)=B_0r^{\theta-2}(1+B_1r^b)\sp C(r)=r^{\theta-2}(1+C_1r^b)\\
& a={4\over \delta-\gamma}\sp \theta={4\d\over \delta-\gamma}\sp z={3\d^2-\g^2+2\g\d-4\over \d^2-\g^2}\sp b=2a(\chi-\g)\sp \beta=2-z+a(\chi-\g)\\
& B_0={(\theta - z - 2) (\theta - z - 1)\over V_0}e^{\d\phi_0}\sp h^2=2{z-1\over 1+z-\theta}{V_0\over Z_0}e^{-(\g+\d)\phi_0}\end{split}\ee

The amplitudes $D_1,B_1,C_1,\Phi_1$ are fixed (after choosing a gauge) and are proportional to $h^2{W_0^2\over Z_0}$, while $A_1$ is proportional to $h{W_0\over Z_0}$. The region of validity of this solution is smaller than that of \ref{IIa} as we also require that $b,\beta$ be subleading in the IR.

\subsubsection{Electric solution at finite charge density}\label{Ia1}
The leading order solution in this case is described by \ref{Ia}.
\be\begin{split} & ds^2=-D(r)dt^2+B(r)dr^2+C(r)dx_idx^i\sp  e^{\phi}=e^{\phi_0+\Phi_1 r^b}r^{a}\sp A_t=\m+A_0r^{\theta-2-z}(1+A_1r^\beta)\\
& D(r)=r^{\theta-2z}(1+D_1r^b)\sp B(r)=B_0r^{\theta-2}(1+B_1r^b)\sp C(r)=r^{\theta-2}(1+C_1r^b)\\
& a={4\over \delta+\gamma}\sp \theta={4\d\over \delta+\gamma}\sp z={3\d^2-\g^2-2\g\d-4\over \d^2-\g^2}\sp b=\beta=a\chi\\
& B_0={(\theta - z - 2) (\theta - z - 1)\over V_0}e^{\d\phi_0}\sp q^2=2{z-1\over 1+z-\theta}{V_0  Z_0}e^{(\g-\d)\phi_0} \sp A_0=\sqrt{2(z-1)\over Z_0 e^{\g\phi_0}(2+z-\theta)}\end{split}\ee

All the amplitudes $D_1,B_1,C_1,\Phi_1,A_1$ are proportional to $hW_0/Z_0$. The region of validity of this solution is smaller than that of \ref{Ia} as we also require that $b,\beta$ be subleading in the IR.

\subsubsection{Magnetic solution at finite charge density}
The leading order solution in this case is described by \ref{IIIa}.
\be\begin{split} & ds^2=-D(r)dt^2+B(r)dr^2+C(r)dx_idx^i\sp  e^{\phi}=e^{\phi_0+\Phi_1 r^b}r^{a}\sp A_t=\m+A_0r^{2-z}(1+A_1r^\beta)\\
& D(r)=r^{4-2z}(1+D_1r^b)\sp B(r)=B_0r^{2}(1+B_1r^b)\sp C(r)=r^{2}(1+C_1r^b)\\
& a={4\over \delta}\sp \theta={4}\sp z={3\d^2-4\over \d^2}\sp b=\beta=a\chi\\
& B_0={(2 - z) (3 - z)\over V_0}e^{\d\phi_0}\sp q^2+h^2Z_0^2=2{z-1\over z-3}{V_0  Z_0}e^{-\d\phi_0} \sp A_0=\sqrt{2(z-1)\over Z_0(z-2)}\end{split}\ee
All the amplitudes $D_1,B_1,C_1,\Phi_1,A_1$ are proportional to $hW_0/Z_0$. The region of validity of this solution is smaller than that of \ref{IIIa} as we also require that $b,\beta$ be subleading in the IR.


\addcontentsline{toc}{section}{References}
\begin{thebibliography}{99}


\bibitem{Mald}
  J. Maldacena,
  {\em ``The Large N Limit of Superconformal field theories and supergravity''}
  \hri{9711200v3}{[hep-th]}.

      \bibitem{witten}
      E. Witten,
      {\em ``Anti de Sitter Space
      and Holography''}
      \hri{9802150v2}{[hep-th]}.


  \bibitem{Singularities}
    S. S. Gubser,
    {\em ``Curvature Singularities: the Good, the Bad, and the Naked''}
    \hri{0002160v1}{[hep-th]}.

  \bibitem{Hartnoll}
  S.~A.~Hartnoll, C.~P.~Herzog and G.~T.~Horowitz,
  {\em ``Building a Holographic Superconductor,''}
  Phys.\ Rev.\ Lett.\  {\bf 101} (2008) 031601
  \hri{0803.3295}{[hep-th]}.

  \bibitem{gk}
  B.~Gouteraux and E.~Kiritsis,
  {\em ``Quantum critical lines in holographic phases with (un)broken symmetry''}
  JHEP {\bf 1304} (2013) 053
  \hri{1212.2625}{[hep-th]}.
  %%CITATION = ARXIV:1212.2625;%%
  %19 citations counted in INSPIRE as of 09 Feb 2014

  \bibitem{st}
    E.~Kiritsis,
    {\em ``String theory in a nutshell,''}
    Princeton University Press, 2007

\bibitem{eht}
  C.~Charmousis, B.~Gouteraux, B.~S.~Kim, E.~Kiritsis and R.~Meyer,
  {\em ``Effective Holographic Theories for low-temperature condensed matter systems,''}
  JHEP {\bf 1011} (2010) 151
    \hri{1005.4690} {[hep-th]}.


   \bibitem{magn}
  E.~D'Hoker and P.~Kraus,
  {\em ``Quantum Criticality via Magnetic Branes,''}
  Lect.\ Notes Phys.\  {\bf 871} (2013) 469
    \hri{1208.1925} {[hep-th]}.

  \bibitem{QCD}
  J. Erlich,
        {\em ``Recent Results in AdS/QCD''}
        \hri{0812.4976v1}{[hep-th]}.

  \bibitem{ADSCMT}
  S. Gubser,
        {\em ``Breaking an Abelian gauge symmetry near a
        black hole horizon''}
        \hri{0801.2977v1}{[hep-th]}.

 \bibitem{gk2}
  B.~Gouteraux, E.~Kiritsis,
        {\em ``Generalized Holographic Quantum Criticality at Finite Density''}
        \hri{1107.2116}{[hep-th]}.
        
\bibitem{dab}
  A.~Dabholkar, R.~Kallosh and A.~Maloney,
  {\em ``A Stringy cloak for a classical singularity,''}
  JHEP {\bf 0412} (2004) 059
  \hri{0410076}{[hep-th]}.
  
  \bibitem{SFI}
  Y. Brihaye, B. Hartmann,
        {\em ``Holographic superfluid/fluid/insulator phase transitions in 2+1
        dimensions,''}
        \hri{1101.5708v1}{[hep-th]}.

\bibitem{QCD0}
  U.~Gursoy, E.~Kiritsis, L.~Mazzanti, G.~Michalogiorgakis and F.~Nitti,
  {\em ``Improved Holographic QCD,''}
  Lect.\ Notes Phys.\  {\bf 828} (2011) 79
    \hri{arXiv:1006.5461}{[hep-th]}.

   \bibitem{QCD1}
     E.~Witten,
     {\em ``Anti-de Sitter space, thermal phase transition, and confinement in gauge theories,''}
     Adv.\ Theor.\ Math.\ Phys.\  {\bf 2} (1998) 505,
    \hri{9803131}{[hep-th]}.


   \bibitem{QCD2}
     T.~Sakai and S.~Sugimoto,
     {\em ``Low energy hadron physics in holographic QCD,''}
     Prog.\ Theor.\ Phys.\  {\bf 113} (2005) 843
     \hri{0412141}{[hep-th]}.

   \bibitem{QCD3}
     J.~Erlich, E.~Katz, D.~T.~Son and M.~A.~Stephanov,
     {\em ``QCD and a holographic model of hadrons,''}
     Phys.\ Rev.\ Lett.\  {\bf 95} (2005) 261602
     \hri{0501128}{[hep-th]}.

   \bibitem{QCD4}
     L.~Da Rold and A.~Pomarol,
     {\em ``Chiral symmetry breaking from five dimensional spaces,''}
     Nucl.\ Phys.\ B {\bf 721} (2005) 79
     \hri{0501218}{[hep-th]}.

   \bibitem{QCD5}
     U.~Gursoy and E.~Kiritsis,
     {\em ``Exploring improved holographic theories for QCD: Part I,''}
     JHEP {\bf 0802} (2008) 032
     \hri{0707.1324}{[hep-th]}.

   \bibitem{QCD6}
     U.~Gursoy, E.~Kiritsis and F.~Nitti,
     {\em ``Exploring improved holographic theories for QCD: Part II,''}
     JHEP {\bf 0802} (2008) 019
        \hri{0707.1349}{[hep-th]}.

   \bibitem{QCD7}
     S.~S.~Gubser and A.~Nellore,
     {\em ``Mimicking the QCD equation of state with a dual black hole,''}
     Phys.\ Rev.\ D {\bf 78} (2008) 086007
           \hri{0804.0434}{[hep-th]}.


   \bibitem{QCD8}
     U.~Gursoy, E.~Kiritsis, L.~Mazzanti and F.~Nitti,
     {\em ``Holography and Thermodynamics of 5D Dilaton-gravity,''}
     JHEP {\bf 0905} (2009) 033           \hri{0812.0792}{[hep-th]}.

   \bibitem{QCD9}
     I.~Iatrakis, E.~Kiritsis and A.~Paredes,
     {\em ``An AdS/QCD model from Sen's tachyon action,''}
     Phys.\ Rev.\ D {\bf 81} (2010) 115004
     \hri{1003.2377}{[hep-th]}.

   \bibitem{QCD10}
     U.~Gursoy, E.~Kiritsis, L.~Mazzanti, G.~Michalogiorgakis and F.~Nitti,
     {\em ``Improved Holographic QCD,''}
     Lect.\ Notes Phys.\  {\bf 828} (2011) 79
     \hri{1006.5461}{[hep-th]}.

   \bibitem{QCD11}
     I.~Iatrakis, E.~Kiritsis and A.~Paredes,
     {\em ``An AdS/QCD model from tachyon condensation: II,''}
     JHEP {\bf 1011} (2010) 123
     \hri{1010.1364}{[hep-th]}.

    \bibitem{QCD12}
      M.~Jarvinen and E.~Kiritsis,
      {\em ``Holographic Models for QCD in the Veneziano Limit,''}
      JHEP {\bf 1203} (2012) 002
     \hri{1112.1261}{[hep-th]}.


\bibitem{FG1}
  S.~Bhattacharyya, S.~Jain, S.~Minwalla and T.~Sharma,
 {\em ``Constraints on Superfluid Hydrodynamics from Equilibrium Partition Functions,''}
  JHEP {\bf 1301} (2013) 040
     \hri{1206.6106}{[hep-th]}.


\bibitem{FG2}
  N.~Banerjee, J.~Bhattacharya, S.~Bhattacharyya, S.~Jain, S.~Minwalla and T.~Sharma,
  {\em ``Constraints on Fluid Dynamics from Equilibrium Partition Functions,''}
  JHEP {\bf 1209} (2012) 046
     \hri{1203.3544}{[hep-th]}.

\bibitem{FG3}
  S.~Minwalla,
   {\em ``Applications of the AdS/CFT correspondence,''}
  Pramana {\bf 79} (2012) 1075.

\bibitem{FG4}
  V.~E.~Hubeny, S.~Minwalla and M.~Rangamani,
   {\em ``The fluid/gravity correspondence,''}
     \hri{1107.5780}{[hep-th]}.


\bibitem{FG5}
  J.~Bhattacharya, S.~Bhattacharyya, S.~Minwalla and A.~Yarom,
   {\em ``A Theory of first order dissipative superfluid dynamics,''}
  JHEP {\bf 1405} (2014) 147
       \hri{1105.3733}{[hep-th]}.


\bibitem{FG6}
  J.~Bhattacharya, S.~Bhattacharyya and S.~Minwalla,
   {\em ``Dissipative Superfluid dynamics from gravity,''}
  JHEP {\bf 1104} (2011) 125
       \hri{1101.3332}{[hep-th]}.

\bibitem{FG7}
  V.~E.~Hubeny, M.~Rangamani, S.~Minwalla and M.~Van Raamsdonk,
   {\em ``The fluid-gravity correspondence: The membrane at the end of the universe,''}
  Int.\ J.\ Mod.\ Phys.\ D {\bf 17} (2009) 2571.

\bibitem{FG8}
  S.~Bhattacharyya, S.~Minwalla and S.~R.~Wadia,
   {\em ``The Incompressible Non-Relativistic Navier-Stokes Equation from Gravity,''}
  JHEP {\bf 0908} (2009) 059
       \hri{0810.1545}{[hep-th]}.

\bibitem{FG9}
  S.~Bhattacharyya, R.~Loganayagam, I.~Mandal, S.~Minwalla and A.~Sharma,
   {\em ``Conformal Nonlinear Fluid Dynamics from Gravity in Arbitrary Dimensions,''}
  JHEP {\bf 0812} (2008) 116
       \hri{0809.4272}{[hep-th]}.

\bibitem{FG10}
  S.~Bhattacharyya, R.~Loganayagam, S.~Minwalla, S.~Nampuri, S.~P.~Trivedi and S.~R.~Wadia,
   {\em ``Forced Fluid Dynamics from Gravity,''}
  JHEP {\bf 0902} (2009) 018
       \hri{0806.0006}{[hep-th]}.

\bibitem{FG11}
  S.~Bhattacharyya, V.~E.~Hubeny, R.~Loganayagam, G.~Mandal, S.~Minwalla, T.~Morita, M.~Rangamani and H.~S.~Reall,
   {\em ``Local Fluid Dynamical Entropy from Gravity,''}
  JHEP {\bf 0806} (2008) 055
       \hri{0803.2526}{[hep-th]}.

\bibitem{FG12}
  S.~Bhattacharyya, V.~E.~Hubeny, S.~Minwalla and M.~Rangamani,
   {\em ``Nonlinear Fluid Dynamics from Gravity,''}
  JHEP {\bf 0802} (2008) 045
       \hri{0712.2456}{[hep-th]}.
       
       
 \bibitem{FG13}
   K.~Jensen, M.~Kaminski, P.~Kovtun, R.~Meyer, A.~Ritz and A.~Yarom,
   {\em ``Towards hydrodynamics without an entropy current,''}
   Phys.\ Rev.\ Lett.\  {\bf 109} (2012) 101601
    \hri{1203.3556}{[hep-th]}.
    
    \bibitem{FG14}
      K.~Jensen, M.~Kaminski, P.~Kovtun, R.~Meyer, A.~Ritz and A.~Yarom,
      {\em ``Parity-Violating Hydrodynamics in 2+1 Dimensions,''}
      JHEP {\bf 1205} (2012) 102
    \hri{1112.4498}{[hep-th]}.     


  \bibitem{SCI}
  T. Nishioka, S. Ryu, T. Takayanagi,
        {\em ``Holographic Superconductor/Insulator Transition
        at Zero Temperature''}
        \hri{0911.0962}{[hep-th]}.


  \bibitem{hooft}
  G. 't Hooft,
        {\em ``A PLANAR DIAGRAM THEORY FOR STRONG INTERACTIONS''}
        Nuclear Physics, B72, (1974), pp. 461-473.

    \bibitem{bek}
    J. Bekenstein,
          {\em ``Black Holes and the Second Law''}
          Lettere al Nuovo Cimento, Vol. 4, No 15 (August 12, 1972), pp. 737-740.

           \bibitem{EMD}
           B. Gouteraux, J. Smolic, M. Smolic, K. Skenderis and M.
           Taylor
           {\em ``Holography for Einstein-Maxwell-dilaton theories from
           generalized dimensional reduction''}
           \hri{1110.2320v2}{[hep-th]}.

   \bibitem{FGC}
   M. Rangamani,
         {\em ``Gravity and Hydrodynamics:
         Lectures on the fluid-gravity correspondence''}
         \hri{0905.4352}{[hep-th]}.

    \bibitem{HSS}
      L.~Huijse, S.~Sachdev and B.~Swingle,
      {\em ``Hidden Fermi surfaces in compressible states of gauge-gravity duality,''}
      Phys.\ Rev.\ B {\bf 85} (2012) 035121
               \hri{1112.0573}{[cond-mat.str-el]}.



    \bibitem{RN1}
    A. Chamblin, R. Emparan, C. V. Johnson, and R. C. Myers,
          {\em ``Charged AdS black holes and
              catastrophic holography''}
          \hri{9902170}{[hep-th]}.


  \bibitem{RN2}
           A. Chamblin, R. Emparan, C. V. Johnson, and R. C. Myers,
                         {\em ``Holography, thermodynamics
                    and fluctuations of charged AdS black holes''}
                    \hri{9904197}{[hep-th]}.

    \bibitem{NG}
    T. Goto,
          {\em ``Relativistic quantum mechanics of one-dimensional mechanical continuum and subsidiary condition of dual resonance model''},
          Progress of Theoretical Physics, Vol. 46, (1971), pp. 1560-1569.

     \bibitem{SC1}
       S.~A.~Hartnoll, C.~P.~Herzog and G.~T.~Horowitz,
       {\em ``Building a Holographic Superconductor,''}
       Phys.\ Rev.\ Lett.\  {\bf 101} (2008) 031601
         \hri{0803.3295}{[hep-th]}.

     \bibitem{SC2}
       S.~A.~Hartnoll, C.~P.~Herzog and G.~T.~Horowitz,
       {\em ``Holographic Superconductors,''}
       JHEP {\bf 0812} (2008) 015, \hri{0810.1563}{[hep-th]}.


     \bibitem{SC3}
       J.~P.~Gauntlett, J.~Sonner and T.~Wiseman,
       {\em ``Holographic superconductivity in M-Theory,''}
       Phys.\ Rev.\ Lett.\  {\bf 103} (2009) 151601
         \hri{0907.3796}{[hep-th]}.

     \bibitem{SC4}
       J.~P.~Gauntlett, J.~Sonner and T.~Wiseman,
       {\em ``Quantum Criticality and Holographic Superconductors in M-theory,''}
       JHEP {\bf 1002} (2010) 060
         \hri{0912.0512}{[hep-th]}.

     \bibitem{SC5}
       S.~Franco, A.~Garcia-Garcia and D.~Rodriguez-Gomez,
       {\em ``A General class of holographic superconductors,''}
       JHEP {\bf 1004} (2010) 092
         \hri{0906.1214}{[hep-th]}.

     \bibitem{SC6}
       S.~S.~Gubser, C.~P.~Herzog, S.~S.~Pufu and T.~Tesileanu,
       {\em ``Superconductors from Superstrings,''}
       Phys.\ Rev.\ Lett.\  {\bf 103} (2009) 141601
          \hri{0907.3510}{[hep-th]}.

     \bibitem{SC7}
       F.~Aprile and J.~G.~Russo,
       {\em ``Models of Holographic superconductivity,''}
       Phys.\ Rev.\ D {\bf 81} (2010) 026009
          \hri{0912.0480}{[hep-th]}.

     \bibitem{SC8}
       Y.~Liu and Y.~W.~Sun,
       {\em ``Holographic Superconductors from Einstein-Maxwell-Dilaton Gravity,''}
       JHEP {\bf 1007} (2010) 099
          \hri{1006.2726}{[hep-th]}.

     \bibitem{SC9}
       A.~Salvio,
       {\em ``Holographic Superfluids and Superconductors in Dilaton-Gravity,''}
       JHEP {\bf 1209} (2012) 134
          \hri{1207.3800}{[hep-th]}.

     \bibitem{SC10}
       S.~S.~Gubser and A.~Nellore,
       {\em ``Ground states of holographic superconductors,''}
       Phys.\ Rev.\ D {\bf 80} (2009) 105007
          \hri{0908.1972}{[hep-th]}.

     \bibitem{SC11}
       G.~T.~Horowitz and M.~M.~Roberts,
       {\em ``Zero Temperature Limit of Holographic Superconductors,''}
       JHEP {\bf 0911} (2009) 015
          \hri{0908.3677}{[hep-th]}.

\bibitem{SC12}
  A.~Donos, B.~Goutéraux and E.~Kiritsis,
       {\em ``Holographic Metals and Insulators with Helical Symmetry,''}
  JHEP {\bf 1409} (2014) 038
  \hri{1406.6351}{[hep-th]}.

\bibitem{SC13}
  A.~Donos and S.~A.~Hartnoll,
  {\em ``Interaction-driven localization in holography,''}
  Nature Phys.\  {\bf 9} (2013) 649
  \hri{1212.2998}{[hep-th]}.
  
\bibitem{SC14}
  A.~Donos and J.~P.~Gauntlett,
  {\em ``Novel metals and insulators from holography,''}
  JHEP {\bf 1406} (2014) 007
  \hri{1401.5077}{[hep-th]}.
  
\bibitem{SC15}
  B.~Goutéraux,
  {\em ``Charge transport in holography with momentum dissipation,''}
  JHEP {\bf 1404} (2014) 181
  \hri{1401.5436}{[hep-th]}.    

  \bibitem{br}
  O. Aharony, S. S. Gubser,
  J. Maldacena,
  H. Ooguri,
  Y. Oz,
  {\em ``Large
  N Field Theories,
  String Theory and Gravity''}
  \hri{9905111v3}{[hep-th]}.



    \bibitem{AdS/CFT}
      A.~V.~Ramallo,
      {\em ``Introduction to the AdS/CFT correspondence,''}
  \hri{1310.4319}{[hep-th]}.

\bibitem{gout}
  B.~Goutéraux,
  {\em ``Universal scaling properties of extremal cohesive holographic phases,''}
  JHEP {\bf 1401} (2014) 080
  \hri{1308.2084}{[hep-th]}.

\bibitem{emad}
  M.~Lippert, R.~Meyer and A.~Taliotis,
  {\em ``A holographic model for the fractional quantum Hall effect,''}
  \hri{1409.1369}{[hep-th]}.

      \bibitem{mag1}
        A.~Almheiri,
        {\em ``Magnetic AdS2 x R2 at Weak and Strong Coupling,''}
          \hri{1112.4820}{[hep-th]}.

      \bibitem{mag2}
        A.~Almuhairi,
        {\em ``AdS$_3$ and AdS$_2$ Magnetic Brane Solutions,''}
          \hri{1011.1266}{[hep-th]}.

      \bibitem{mag3}
        E.~D'Hoker and P.~Kraus,
        {\em ``Magnetic Brane Solutions in AdS,''}
        JHEP {\bf 0910} (2009) 088
          \hri{0908.3875}{[hep-th]}.

      \bibitem{mag4}
        E.~D'Hoker and P.~Kraus,
        {\em ``Charged Magnetic Brane Solutions in AdS (5) and the fate of the third law of thermodynamics,''}
        JHEP {\bf 1003} (2010) 095
          \hri{0911.4518}{[hep-th]}.

      \bibitem{mag5}
        E.~D'Hoker and P.~Kraus,
        {\em ``Holographic Metamagnetism, Quantum Criticality, and Crossover Behavior,''}
        JHEP {\bf 1005} (2010) 083
          \hri{1003.1302}{[hep-th]}.

      \bibitem{mag6}
        E.~D'Hoker and P.~Kraus,
        {\em ``Magnetic Field Induced Quantum Criticality via new Asymptotically AdS$_5$ Solutions,''}
        Class.\ Quant.\ Grav.\  {\bf 27} (2010) 215022
          \hri{1006.2573}{[hep-th]}.

      \bibitem{mag7}
        E.~D'Hoker and P.~Kraus,
        {\em ``Charge Expulsion from Black Brane Horizons, and Holographic Quantum Criticality in the Plane,''}
        JHEP {\bf 1209} (2012) 105
          \hri{1202.2085}{[hep-th]}.

      \bibitem{mag8}
        A.~Almuhairi and J.~Polchinski,
        {\em ``Magnetic AdS x R$^2$: Supersymmetry and stability,''}
          \hri{1108.1213}{[hep-th]}.


        \bibitem{Gur}
          U.~Gursoy,
          {\em ``Continuous Hawking-Page transitions in Einstein-scalar gravity,''}
          JHEP {\bf 1101} (2011) 086
            \hri{1007.0500}{[hep-th]}.
\end{thebibliography}
\end{document}